\documentclass[12pt]{article}

\usepackage{amsmath,amsfonts,epsfig,color,latexsym}

\newcommand{\be}{\begin{equation}}
\newcommand{\ee}{\end{equation}}
\newcommand{\bea}{\begin{eqnarray}}
\newcommand{\eea}{\end{eqnarray}}

\renewcommand{\theequation}{\thesection.\arabic{equation}}

\let\newsection=\section
\renewcommand{\section}{\setcounter{equation}{0}\newsection}

\begin{document}

\begin{flushright}
hep-th/0306269\\
BROWN-HET-xxxx
\end{flushright}
\vskip.5in

\begin{center}

{\LARGE\bf Towards a Chern-Simons M theory of $OSp(1|32)\times OSp(1|32)$}
\vskip 1in
\centerline{\Large Horatiu Nastase}
\vskip .5in

\end{center}
\centerline{\large Brown University}
\centerline{\large Providence, RI, 02912, USA}

\vskip 1in

\begin{abstract}

{\large A possible way of defining M theory as the CS theory for the supergroup
$OSp(1|32)\times OSp(1|32)$ is investigated, based on the approach by Horava
in hep-th/9712130. In the high energy limit (expansion in M), where only the
highest ($R^5$) terms survive in the action, the supergroup contracts to the 
D'Auria-Fre M theory supergroup. Then the contracted equations of motion
are solved by the usual 11d supergravity equations of motion, linearized in 
everything but the vielbein. These two facts suggest 
that the whole nonlinear 11d sugra should be obtainable somehow in the 
contraction limit. Type IIB also arises as a contraction of 
the $OSp(1|32)\times OSp(1|32)$ theory. The presence of a cosmological constant
in 11d constraints the parameter M experimentally to be of the order of 
the inverse
horizon size, $1/L_0$. Then the 11d Planck mass $M_{P,11}\sim 10GeV$ 
(hopefully higher: $>TeV$ due to uncertainties). Unfortunately, the most naive 
attempt at 
cosmological implications for the theory is excluded experimentally. 
Interestingly, the low energy expansion (high M) of the CS theory, truncated
to the gravitational sector, gives much better phenomenology.
}

\end{abstract}

\newpage

\section{Introduction}

One of the most important theoretical problems nowadays is finding a way to 
naturally accomodate a cosmological constant in string and M theory. 
In string theory one can add a mass term to IIA supergravity, but that does 
not generate a $\Lambda$. Progress towards the embedding of massive IIA 
theory in M theory has be made in \cite{hull,lnr}. But no one has been able to 
write down a consistent theory extending the usual 11d supergravity 
\cite{cjs} and 
incorporating a mass term \cite{cfgpv}.
No-go theorems \cite{bdhs} always assume that the 
field content is the same as in usual 11d supergravity (with possibly 
trading the 3-form for its dual 6-form), and that the low energy action 
is quadratic. 

There is one approach which constructs supergravity theories with cosmological 
constant in any odd dimension. That is the approach of geometric, Chern-Simons 
type supergravities \cite{pvn,at}
The approach has become more than academic with the 
realization, due to Witten \cite{witten}, 
that in 3d, usual gravity (with or without a 
cosmological constant) is the same as CS gravity theory, and consequently 
can be quantized and renormalized. CS supergravity theories have been written 
in 3d as well \cite{pvn,at}. 
Going to higher dimensions, one encounters a problem though:
a CS-type action will be dominated either by the cosmological term (lowest)
or by an $R^n$ term (highest). In 5 dimensions, one can dimensionally reduce 
to a good 4d Einstein theory \cite{chamseddine}.
But in higher dimensions, that avenue isn't 
available either.

In eleven dimensions, the choice of possible supergroup for a CS theory 
usually revolves around $OSp(1|32)$, since the original paper \cite{cjs}.
Then, D'Auria and Fre \cite{df} 
wrote an (almost) geometric formulation of the 11d 
supergravity based on a supergroup involving an extra spinor and 
5-form gauge fields.   
In \cite{btz,tz1,tz2,zanelli} a CS theory of the $OSp(1|32)$ supergroup 
was found, but no relation to M theory was claimed. 
In a paralel effort, Horava \cite{horava}
tried actually to connect a CS theory of the 
supergroup $OSp(1|32)\times OSp(1|32)$ to M theory. However, his claim to 
obtain 11d supergravity as a low energy theory seems hard to support 
given that the supergroup contracts to the Poincare group in the high 
energy limit. Finally, in \cite{banados}, one tried to linearize the 
$OSp(1|32)$ equations of motion in the high energy limit, 
with the claim that the 11d supergravity linearized equations of motion 
arise. I will actually show that is not the case, and the approximation 
used in \cite{banados} is invalid, yet in the low energy limit, one can 
still get close to 11d supergravity. 

Note that a 11d CS supergravity can 
be written as a topological theory in (10,2) dimensions. Previous approaches 
at constructing a 
usual (nontopological) (10,2) supergravity failed \cite{cfgpv2}.

The subject of this paper is to connect these approaches. I will first 
start with the $OSp(1|32) \times OSp(1|32)$ supergroup and show explicitly 
the (almost) contraction to the D'Auria-Fre supergroup. The almost part 
is a mismatch in numbers and is attributed to the auxiliary 0-form 
introduced by D'Auria and Fre. I will then use this contraction to construct 
an action for the supergroup and define the contraction on the action. 
Type IIB can also be obtained from the CS action written in (10,2) notation. 
I will analyze the equations of motion of the CS theory and will prove the 
fact that the equations of motion of usual 11d sugra, linearized in everything 
but the vielbein, solve the CS equations of motion in the high energy limit, 
as expected from the fact that the supergroup contracts in the right way. 

I will then analyze the possible phenomenological implications of having 
a CS M theory. I will define the Planck mass and find out a constraint on 
the total volume in the Universe. 
An observation of Horava was that in a CS theory one 
can introduce matter as Wilson lines, and in a mean field approximation 
this produces a cosmological constant term of opposite sign (dS). So 
if they mismatch, one can have even a de Sitter background.
The fact that one has a cosmological 
constant from the start means that there are experimental constraints. It 
will turn out that the expansion parameter M is constrained to be of the 
order of the horizon size, $M\sim 1/L_0$. In turn, one can construct the 
simplest of cosmological models; but we will find that it is excluded. 

The possible embedding of M theory in the CS sugra has however the problem 
that it is not clear how to quantize the system. I will present a few 
ideas, but the basic problem is that the dominant term is not the 
quadratic kinetic term, so the usual perturbative expansion fails. 

The last question to be addressed is what happens in the low energy limit.
I show that the 11d sugra equations of motion are not satisfied because of  
mass terms. However, if one restricts to the gravitational action, one seems 
to get a consistent theory, and analyze the resulting 
cosmology, which is much 
better than in the high energy limit case. 

I have tried to be as self-contained as possible, so the paper is somewhat 
long, but it shouldn't be too hard to read.

The paper is organized as follows. In section 2 I present a review of CS 
supergravities, for completeness, since it is a subject which is not very 
familiar. In section 3 I review the arguments for $OSp(1|32)$ 
invariance in M theory and the d'Auria-Fre formulation. In section 4 I prove
the contraction of $OSp(1|32)\times OSp(1|32)$ to the D'Auria-Fre supergroup 
and derive the action. In section 5 I prove the contraction of the CS model 
to IIB, at the level of supergroups.
In section 6 I study the equations of motion in 
the ``high energy limit'', and in section 7 I try to define the limit better.
Then in section 8 I study the phenomenological consequences and the 
cosmological constant. In section 9 I present ideas about the quantum theory,
and in section 10 I analyze the ``low energy limit'' and its phenomenological 
consequences. I finish in section 11 with conclusions.

\section{Chern-Simons supergravities}

In this section I will quickly
review the subject of Chern-Simons (super)gravities from the perspective 
of what we will need later on. 

The problem of writing gravity as a Chern Simons theory of the Poincare 
group has been around for quite some time. The idea would be that 
the spin connection $\omega^{ab}_{\mu}$ and the vielbein $e^a_{\mu}$ are 
viewed as gauge fields in a space with no metric (the action is topological,
so does not involve the metric for contraction; one could think of the 
vielbein as a gauge field in an auxiliary space) and one tries to derive 
Einstein gravity as a Yang-Mills type action. 

In 4 dimensions, since the curvature 2-form is 
\be
R^{ab}=2[d\omega^{ab}+\omega^{ac}\wedge \omega^{cb}]
\ee
the Einstein action is 
\be
S_{4d, EH}(e, \omega )=\int R^{ab}\wedge e^c\wedge e^d \epsilon_{abcd}
\ee
Note that we are in a first order formulation, since $\omega$ and $e$ 
are considered independent gauge fields.
Varying with respect to $\omega$ one gets the vielbein constraint $De=0$. Why
isn't it a Yang-Mills action for the Poincare group (with gauge fields
$\omega^{ab}$ and $e^a$, corresponding to the generators $J^{ab}$ and 
$P^a$)?

There are two points one should make:
a) The action is of the type 
\be
{\cal L} =2 (d\omega +\omega \wedge \omega)\wedge e \wedge e \sim (dA +A\wedge
A)\wedge A \wedge A
\ee
and so is not in a gauge invariant form. However,

b) it is gauge invariant under the Poincare group gauge transformation
$\delta \omega^{ab}=0, \delta e^a =(D\lambda)^a
=d\lambda^a +\omega^{ab}\wedge \lambda^b$, 
but only on-shell: if $De^a=0$ (no torsion) and one correspondingly uses 
$e^{-1}$ to identify local  translations in the base manifold 
(diffeomorphisms with parameter $\lambda^{\mu}$) with local 
translations in the tangent space (gauge transformations), by
 $\lambda^{\mu}=e^{\mu}_a \lambda^a$. 

In three dimensions however, the Einstein action
\be
{\cal L}= 2(d\omega +\omega \wedge \omega)\wedge e \sim dA\wedge A
+\frac{2}{3} A\wedge A \wedge A 
\ee
is gauge invariant, being of Chern-Simons type \cite{witten}. 
Indeed, $d{\cal L}=F\wedge
F$, so the Poincare group ISO(d-1,1) has gauge field 
\be
A_{\mu}=e_{\mu}^a P_a +\omega^{ab}_{\mu} J_{ab}
\ee
A Chern-Simons action is well defined once you give a prescription for 
the trace over the gauge group indices that is, once you give a group 
invariant form. The problem of group invariants is very difficult in general, 
as there doesn't seem to be an algorithmic way of finding them. 

In a general dimension d,
  $W_1=x J_{ab}J^{ab}+y P_a P^a$ is invariant only if x=0, so 
is degenerate. But in 3d one also has $W_2=\epsilon _{abc} J^{ab}P^c$,
corresponding to the bilinear
form $d^{AB}\sim <P^a J^{bc}>=\epsilon_{abc}$, 
or defining $J^a=1/2 \epsilon^{abc}J_{bc}$, we have $<J_a, P_b>=\delta
_{ab}, <J_a, J_b>=<P_a, P_b>=0$. 

Then the ISO(2,1)-invariant CS action is 
\be
S_{CS}=\int_{M_4} F^A\wedge F^B d_{AB}=\int _{M_3}(dA \wedge A + \frac{2}{3}
A\wedge A \wedge A )^{AB}d_{AB}
\ee
Can also be extended to the dS/AdS cases as well, by 
\be
S=S_{EH}+\lambda \int e\wedge e \wedge e
\ee
which is invariant under SO(3,1) or SO(2,2) depending on the sign 
of $\lambda$. It can also be extended to 3d supergravity. 

One would like to do something similar in other dimensions and for
other models. 

The idea is known as ``gauging superalgebras''. E.g., \cite{pvn} did this for 
conformal supergravity in 3d, with gauge group $OSp(1|4)$, based on the
work in 4d \cite{ktv}. The CS-type
action is 
\be
S_{CS}(4d)=\int _{M_4 }\gamma_{AB} R^A \wedge R^B =\int _{M_3}
(\gamma_{AB} R^A \omega ^B +\frac{1}{6} f_{ABC}\omega ^C \omega^B
\omega^A)
\ee
and is $OSp(1|4)$ invariant as it stands, but then the Q-gauge
transformation is not a local susy. One can see that then $\{ Q, Q \}
\omega^{ab}\sim \delta_P \omega^{ab}=0$, which is clearly not desired 
(we should have a general coordinate transformation on the r.h.s.).
One needs to impose
constraints in order to have a conformal sugra action. The needed
constraints turn out to be (see also \cite{pvn2} for more details)
\be
R_{\mu\nu}^a(P)=0,\;\; R_{\mu\nu}^{\alpha}(Q)=0, \;\; R_{\mu\nu}^{ab}(M)=0
\ee
where $R_{\mu\nu}^a(P)=T_{\mu\nu}^a+$ fermions. 
In general, also local translations ($P_m$ gauge transformations)
are not general coordinate transformations, i.e. gravity is not a 
local gauge theory if the Poincare group.  The local super-Poincare 
algebra is different than the global one! Even in 3d, the general 
coordinate transformations differ from the gauge transformations by 
equation of motion terms ($R_{\mu\nu}^a(P)$ and $R_{\mu\nu}^{ab}(M)$).

However, we will ignore this fact in the rest of the paper 
and just concentrate 
on Chern-Simons actions and the solutions to their equation of motion
in the absence of constraints. The solutions 
we will be looking for in the follwing will often amount to existence 
of constraints, most notably the no-torsion contraint, $R_{\mu\nu}^m(P)=0$.

Chamseddine \cite{chamseddine} 
noticed that in 2n+1 dimensions there is the Chern-Simons
gravitational action
\be
{\cal S}_{2n+1}=k \int _{M_{2n+1}}\omega_{2n+1}, \;\; 
\omega_{2n+1}=(m+1)\int _0^1 dt <A (tdA +t^2 A^2)^n>
\ee
which is contracted with the n+1 dimensional SO(2n+2) invariant 

\be
<J_{a_1 b_1}...J_{a_{n+1}b_{n+1}}>=\epsilon_{a_1b_1...a_{n+1}b_{n+1}}
\ee
for dS (SO(2n+1,1)) or AdS (SO(2n,2)) 
invariance groups. Here $J_{2n+1 a}\equiv P_a$ and the 
gauge fields are $A^{ab}=\omega^{ab}, A^{a, 2n+1}=e^a, a=0,...,2n$. 

This implies that 
\bea
&&{\cal S}_{2n+1}=l\int_{M_{2n+1}} \epsilon_{a_1...a_{2n+1}}\sum
_{l=0}^n \frac{\lambda^l}{(2l+1)}\begin{pmatrix} l&\\n&\end{pmatrix}
R^{a_1a_2}\wedge ...\nonumber\\&&
\wedge R^{a_{2n-2l-1}, a_{2n-2l}}\wedge
e^{a_{2n-2l+1}}\wedge ... \wedge e^{a_{2n+1}}
\eea
and where 
\be
\lambda=- {\rm for} \;\; SO(1,2n+1), \; + {\rm for} \;\; SO(2,2n) \;
{\rm and} \;\; 0 \; {\rm for} ISO(1,2n)
\ee
and $R^{ab}=d\omega^{ab}+\omega^{ac}\wedge \omega^{cb}$. We notice 
that the Einstein-Hilbert action appears only in the (anti) de Sitter case 
since the Poincare group ISO(2n,1) is a Wigner-Inonu contraction 
of the (A)dS group. This will be important later on.

In 5 dimensions, the CS theory is also special. The action is
\be
{\cal S}_{5, dS}=k \int_{M_5}(e\wedge R\wedge R+2/3 \lambda e\wedge 
e \wedge e \wedge R +1/5 \lambda^2 e^{\wedge 5})
\ee
The equations of motion for a gravitational CS action in 2n+1 dimensions
are, schematically
\be
\epsilon F_1...F_n=0, F=dA, A=(\omega, e)
\ee
(there is a free index on the epsilon symbol, where we varied the
gauge field) which then give 
\bea
&& \epsilon_{a_1...a_{2n+1}}(R^{a_1a_2}+\lambda e^{a_1}\wedge e^{a_2})
...()^{a_{2n-1}, a_{2n}}=0\nonumber\\
&&\epsilon_{a_1...a_{2n+1}}(R^{a_1a_2}+\lambda e^{a_1}\wedge e^{a_2})
...()^{a_{2n-3}, a_{2n-2}}T^{a_{2n-1}}=0\nonumber\\
&& T^a\equiv De^a= de^a + \omega^{ab} e^b
\label{five}
\eea
Note that $R^{ab}+\lambda e^a\wedge e^b =0, T^a=0$ is a solution of these 
equations independently of dimension, so classically at least EH gravity 
is embeddable in CS gravity.
By expanding around the classical background 
\bea
&& e^a=e_0^a+\bar{e}^a\nonumber\\
&& \omega^{ab}=\omega_0^{ab}+\bar{\omega}^{ab}
\eea
one gets the quadratic lagrangeian in 5d (after a few manipulations)
\be
{\cal L}_{5d, quadr}=k \epsilon_{abcde}[(R_0^{ab}+\lambda e_0^a
e_0^b)(2\bar{e}^c
D_0\bar{\omega}^{de}+\bar{\omega}^{cd}\bar{\omega}^{ef}e_{0,f})
+T_0^a \bar{\omega}^{bc}D_0\bar{\omega}^{de}]
\ee
where
\be
D_0\bar{\omega}^{ab}=d\bar{\omega}^{ab} +\omega_0^{ac}\bar{\omega}^{cb}
+\bar{\omega}^{ac}\omega_{0}^{cb}
\ee
Notice then that if both $(R_0^{ab}+\lambda e_0^ae_0^b)$ and $T^a$ are
nonzero, both $\bar{\omega}$ and $\bar{e}$ propagate. If $T_0^a=0$
one has a metric theory, because then $\omega=\bar{\omega}$ is a
function of e, and if both are zero, one has no quadratic lagrangian. 
But a classical background
doesn't necessarily have to have both of them zero to satisfy the
equations of motion. 

In particular, a good background (amounting to a dimensional reduction) is 
(as one can easily check from the equation (\ref{five}))
\bea
&&T_0^a=0, a=0,4 \nonumber\\
&& R_0^{\alpha \beta}+\lambda e_0^{\alpha} e_0^{\beta}=0, \alpha =0,3
\nonumber\\
&&R_0^{\alpha 4}+\lambda e_0^{\alpha} e_0^{4}\neq 0
\eea
solved by
\bea
&& e_{0, \mu}^{\alpha}=\delta_{\mu}^{\alpha}\frac{1}{1-1/4 \lambda
  x^{\alpha} x_{\alpha}}, \;\; e_{0 4}^4=c ({\rm const})\nonumber\\
&& \omega_{0, \mu}^{\alpha \beta}=-\frac{\lambda}{2}\frac{\delta_{\mu}
^{\alpha}x^{\beta}-\delta_{\mu}^{\beta}x^{\alpha}}{1-1/4 \lambda
x^{\alpha }x_{\alpha}}
\eea
and all other components of $\omega$ and e are zero. Then
$(R_0^{\alpha 4}+\lambda e_0^{\alpha}e_0^4)_{\mu 4}=\lambda c 
\delta_{\mu}^{\alpha}/(1-x^{\alpha}x_{\alpha}/4)$ 
is the only nonzero background component and 
the quadratic lagrangeian is 
\bea
&& {\cal L}= \frac{2 k \lambda c}{1-1/4 x^{\alpha}x_{\alpha}}
[\bar{e}_{\mu}^{\alpha} D_{0\nu}\bar{\omega}_{\rho}^{\beta \gamma}
\delta_{\alpha \beta \gamma}^{\mu\nu \rho}\nonumber\\
&&-\frac{1}{1-1/4 \lambda 
x^{\alpha}x_{\alpha}}(\bar{\omega}_{\mu\nu\rho}\bar{\omega}^{\mu\nu\rho}
-\bar{\omega}_{\mu \rho}^{\mu}\bar{\omega}_{\nu \rho}^{\nu}]
\eea
and then both $(e_{0,\mu}^{\alpha}, \omega_{0, \mu}^{\alpha \beta})$ 
and the quadratic action are the same as for the EH action with
cosmological constant in 4d. 
\be
{\cal S}_4=\frac{k \lambda c}{4}\int \epsilon^{\mu\nu\rho\sigma}
\epsilon_{\alpha \beta\gamma \delta }e_{\mu}^{\alpha} e_{\nu}^{\beta}
(R_{\rho\sigma}^{\gamma\delta}+\frac{1}{2}
\lambda e_{\rho}^{\gamma}e_{\sigma }^{\delta})
\ee
But this is so since the 4d CS-like action (Born-Infeld) obtained 
by dimensionally reducing 5d CS via $e^4_{04}=c, T_0^a=0$ is
\be
{\cal S}_4=A\int (\frac{1}{\lambda}
R\wedge R +2R\wedge e \wedge e +\lambda e\wedge e\wedge
e \wedge e)=\frac{A}{\lambda}\int \sqrt{ det (R^{ab}+\lambda e^a e^b)} 
\ee
and the first term is topological. So really, the CS action dimensionally 
reduces to 4d to a usual action with a topological term, equivalent to a 
BI action. 

Dimensional reduction has a peculiar meaning in the context of CS gravity.
The point is that gravity itself is part of the gauge fields, but the action 
is topological (in particular the inverse metric and the star operation are 
not defined), so it could as well be defined on an auxiliary space with 
a different metric. But by analogy with usual circle dimensional reduction
one can guess what it is. One can choose the gauge $e_4^{\alpha}=0$, then 
the field $e_{\mu}^4dx^{\mu}$ is a gauge field, which could be put to zero, 
and $e^4_4$ is a scalar field, which can be put to a constant.

The above case of Chamseddine reduction is a particular case of a more general 
procedure. There exists a class of gravitational lagrangians with 
particular properties called Lanczos-Lovelock Lagrangeians (LL) 
\cite{lanczos,lovelock,zanelli}
\be
S_G=\int \sum_{p=0}^{[D/2]}\alpha _p {\cal L}^p
\ee
with $\alpha_p$ arbitrary constants and 
\be
{\cal L}_G^p=\epsilon_{a_1...a_D}R^{a_1a_2}...R^{a_{2p-1}a_{2p}}e^{a_{2p+1}}
..e^{a_D}
\ee
They can be understood as the dimensional continuation of Euler densities
(topological invariants) in 2p dimensions and 
have a special significance in string theory. Zwiebach \cite{zwiebach}
showed that the 
most general ghost-free combination of $R^2$ terms in any dimension is 
(one needs to cancel terms with 2 h's and 4 derivatives)
\be
\int \sqrt{g}d^D x y^4=\int \sqrt{g}d ^D x (R^{\mu\nu\rho\sigma}
R_{\mu\nu\rho\sigma}-4R^{\mu\nu}R_{\mu\nu}+R^2)
\ee
and is a total derivative to order $h^2$ in any dimension and for 
$D>4$ describes nontrivial $h^n$ vertices $n\ge 3$. The 3-point vertex
of that combination is what one gets from the 3 on-shell scattering 
of gravitons in bosonic string theory (Virasoro-Shapiro model). 
It is also the term $L^2$ in the LL Lagrangian, i.e. the dimensional
continuation of the 4d Euler density. The EH action is the
dimensional continuation of the 2d Euler density $L^1$, so Zwiebach 
conjectured that the series continues and we have a LL Lagrangian 
$L^1+L^2+L^3+...$ coming from string theory. Each of these terms has
the property that the leading term vanishes: no tadpole around
Minkowski space since there is no linear term in $L^1$, no ghosts
because there is no quadratic term in $L^2$, etc. 
Moreover, the field equations are second order in derivatives 
\cite{lovelock,zumino}.
In fact, the LL lagrangian can be defined as the most general
lagrangian which in the metric form ($T^a=0$) gives field equations 
of second order in derivatives. Unfortunately, Zwiebach's conjecture is 
not correct. Bosonic string theory has a non-LL term at $R^3$ order, and 
superstring theory at order $R^4$ \cite{mt}. More on that later.

There are two important classes of LL lagrangians \cite{zanelli}.
In general,  from the LL lagrangian one can get new
constraints from the equations of motion if one acts with covariant
derivatives. In two cases, that is not so. Let's redefine
$\alpha_p\rightarrow \alpha_p l^{d-2p}$, and now $\alpha_p$ is
dimensionless.

In d=2n+1 we can define CS gravitational actions, in particular
 AdS-CS ones, with 
\be
\alpha '_p=\frac{k}{d-2p}\begin{pmatrix} & n \\& p\end{pmatrix}, 
0\leq p \leq n 
\ee

In d=2n we can define the Born-Infeld (BI) gravitational action:
\be
\alpha_p =k \begin{pmatrix} & n \\& p\end{pmatrix} , 0\leq p \leq n
\ee
The lagrangian is then of BI type, being the pfaffian,
\bea
&&{\cal L}=k \epsilon_{a_1...a_d}(R^{a_1 a_2}+\frac{1}{l^2} e^{a_1}e^{a_2})...
(R^{a_{d-1}a_d}+\frac{1}{l^2}e^{a_{d-1}a_d})\nonumber\\ &&
= k \times pf [R^{ab}+\frac{1}{l^2}e^ae^b]=k \sqrt{det 
[R^{ab}+\frac{1}{l^2}e^ae^b]}
\eea
By dimensional reduction from d=2n+1 of the AdS-CS action
(in the sense defined above for the 5d to 4d 
reduction) one can obtain the d=2n BI case, since one isolates
$e^{2n+1}$ in the CS form, and as out of d places for the 2n+1'th 
d-2p are favourable (on an e, not an R), one gets the coefficient 
\be
\alpha ' _p \frac{d-2p}{d}=\frac{1}{d} \alpha_p
\ee

\section{Arguments for $OSp(1|32)$ symmetry}

Cremmer, Julia and Scherk \cite{cjs}
observed that the group $OSp(1|32)$ is  
the minimal grading of the Sp(32) group, which is the maximal bosonic
group preserving the Majorana property of a SO(10,1) spinor, and is
therefore a natural candidate for an invariance  group of the 11d
sugra. Under $SO(10,1)\in OSp(1|32)$, the generators of $OSp(1|32)$ 
split into $(P^A, J^{AB}, Z_5^{A_1...A_5}, Q_{\alpha} )$. 
van Holten and van Proeyen \cite{vv}
also notice that $OSp(1|32)$ is the minimal grading 
of the AdS algebra SO(10,2) as well. 

So the question appears whether 11d sugra is a gauging of $OSp(1|32)$? 
That was investigated by d'Auria and Fre \cite{df}. 
One could hope that the 7-form
\be
F_7=dB^{A_1...A_5}\wedge e^{A_1}\wedge ... \wedge e^{A_5}
\ee
is the dual of $F_4=dA_3$ present in the 11d sugra. But then the
curvature of the B gauge field 
\be
R^{A_1...A_5}= {\cal
  D}B^{A_1...A_5}-\frac{i}{2}\bar{\psi}\Gamma^{A_1...A_5}\psi
-\alpha_5 \epsilon_{A_1...A_5 BCD EFG}B_{BCD HI}B_{EFG HI}
\ee
is nonabelian, violating the Coleman-Mandula theorem (the B field 
symmetry will be a nonabelian internal symmetry commuting with the 
gauge symmetry of the theory). There is a Wigner-Inonu contraction 
of the theory which makes the curvature abelian, but it turns out that 
it doesn't do the job of reproducing the 11d sugra.

The problem can be traced to the fact that $A_{\mu\nu\rho}$ in 11d
sugra  is a 3-form, not a 1-form. So either $A_3$ is a composite of 
1-forms, or one needs a n-form generalization of Yang-Mills theory. 
The latter was introduced in \cite{df} and named 
Cartan integrable systems (CIS). The
gauge potentials $\Pi^M(p)$ have  curvatures
\be
R^{M(p+1)}=d\Pi^{M(p)}+\sum_{n=1}^N
\frac{1}{n}C^{M(p)}_{N_1(p_1)...N_n(p_n)}\Pi^{N_1(p_1)}...\Pi^{N_n(p_n)}
\ee
d'Auria and Fre found a CIS with 1-forms ($\omega^{AB}, e^A, \psi$), 
corresponding respectively to ($J_{AB},$ $ P_A,$ $ Q$), together with the 
3-form A and the 0-form $F_{A_1...A_4}$, needed in order to give a first order
formulation of 
\be
R^{\Box}=dA-\frac{1}{2}\bar{\psi}\wedge F^{AB}\psi \wedge e_A \wedge e_B
\ee
The action is (schematically; the dots stand for gauge potential 
terms and the numerical coefficients are ignored)
\bea
I&\sim&\int (R^{AB}\wedge ...+R^A\wedge ... +\bar{\rho}\wedge ... +
R^{\Box}\wedge ...)\nonumber\\
&&+\int R^{\Box}\wedge R^{\Box}\wedge A - F_{A_1...
A_4}F^{A_1...A_4}e^{\wedge 11}+F_{A_1...A_4}R^{\Box}\wedge e^{\wedge
7}
\eea
It is not a completely geometric lagrangeian, because of the 0-form. 
The problem is traced to the impossibility of writing the 
3-form kinetic term $\int (dA_3) * (dA_3)$ as a geometric object. 
Geometric objects here means ``topological'', i.e. written without 
reference to the inverse metric or the Hodge star.

They also found a supergroup formulation which is equivalent 
to the CIS. One trades the 3-form A for the 1-forms $B^{AB},
B^{A_1... A_5}, \eta$ (extra 32-component spinor). The ``linear''
piece of the 3-form (linear in fields other than the vielbein) is 
\be
A= B^{AB}\wedge e^A \wedge e^B +...
\ee
The gauge field is then 
\be
\omega^{AB}J_{AB} + e^A P_A +\psi Q +\eta Q' +B^{AB} Z_{AB} 
+B^{A_1...A_5}Z_{A_1...A_5}
\ee
and the remarcable fact is that the superalgebra is the M theory
algebra (with $Z_{AB}$ and $Z_{A_1...A_5}$ identified with the 2-form 
and 5-form central charges of M theory), together with the spinor
charge Q', which modifies the algebra as follows (the two values in 
brackets correspond to two possible supergroups)
\bea
&&[Q, P_A] = i \begin{pmatrix} 1&\\0&\end{pmatrix} \Gamma_A Q'
\nonumber\\
&&[Q, Z_{A_1A_2}] = \begin{pmatrix} 1/5 &\\ -1/2 & \end{pmatrix}
\Gamma_{A_1A_2}Q' \nonumber\\
&&[Q, Z_{A_1...A_5}] = \begin{pmatrix} \frac{1}{240} &\\
  -\frac{1}{144} & \end{pmatrix} \Gamma_{A_1...A_5} Q'
\eea
so the algebra is the M theory algebra if Q' is Wigner-Inonu-contracted 
away ($Q'\rightarrow a Q'\rightarrow 0$). However, it is still not a 
completely geometric formulation, since one still has the 0-form
$F_{A_1...A_4}$.

\section{$OSp(1|32)\times OSp(1|32)$ and action}

Horava \cite{horava}
tried also to use the supergroup approach to define 11d sugra,
and he  noticed that $S_{CS}$ for $OSp(1|32)$ is not parity invariant. 
A possible remedy would be to modify the parity action as 
 ${\cal P}={\cal P}_0 {\cal I}$, where ${\cal I}$ is an action on the 
group corresponding to the action in spacetime, i.e. $P_A, J_{AB}, 
Z_{A_1...A_5}$ to change sign when $A_i=1$. But this is not compatible 
with the Lie algebra. The only way that can be made to work is to make 
duals of everything, and go to the group $OSp(1|32)\times OSp(1|32)$.
This is an important point, since one needs M theory to be
nonperturbatively parity invariant, in order to get the Horava-Witten 
construction and heterotic string theory \cite{hw}. Horava then claimed that 
the $OSp(1|32) \times OSp(1|32)$ group contracts to the d'Auria-Fre group. 
In this section I will try to verify this claim and derive a CS action.

Then the generators are 
\bea
&&P_A, J_{AB}, Z_{A_1...A_5}, Q \;\;{\rm together \;\; with }\nonumber\\
&&Z_A, Z_{AB}, Z'_{A_1...A_5}. Q'{\rm or \;\; rather\;\; their 
\;\; duals }\;\;Z_{(6)}, Z_{(9)}, Z_{(10)}
\eea

One expect then to have among other relations the physical commutation 
relations
\bea
&&\{ Q , Q \} =\Gamma^A P_A +\frac{1}{2}\Gamma^{AB}J_{AB} +\frac{1}{5!}
\Gamma^{A_1...A_5}Z_{A_1...A_5}\nonumber\\
&& [P_A, P_B ] =M^2 J_{AB} \nonumber\\
&&[Q, P_A]=\Gamma_A Q'
\eea
Let us look at the $OSp(1|32)$ algebra. It can be defined as \cite{vv,vp} 
\bea
&&\{Q_{\alpha}, Q_{\beta}\}=M_{\alpha\beta}\nonumber\\
&&[M_{\alpha\beta}, Q_{\gamma}]=Q_{(\alpha}C_{\beta ) \gamma}\nonumber\\
&&[M_{\alpha \beta}, M_{\gamma\delta}]=C_{\alpha (\gamma} M_{\delta )
\beta}+C_{\beta(\gamma}M_{\delta)\alpha}
\eea
where $M_{\alpha\beta}$ is a symmetric 32 by 32 matrix. On that space 
we have the completeness relation (Fierz identity)
\be
\sum_{k=1,2,5} \frac{(-)^{k+1}}{k!}(\Gamma^k)_{\alpha\beta}(\Gamma_k)_{\gamma
\delta}=32 C_{\alpha(\delta} C_{\gamma)\beta}
\ee
where $\Gamma^k$ is a shorthand for $\Gamma^{A_1...A_k}$. So 
$\Gamma^{A}, \Gamma^{AB}$ and $\Gamma^{A_1...A_5}$ form a 
basis. Then writing
\be
M_{\alpha\beta}=\sum_{k=1,2,5}(\Gamma^k)_{\alpha\beta}\frac{1}{k!}
Z_k
\ee
with inverse
\be
Z^j=\frac{(-)^{j+1}}{32}Tr(\Gamma^jM)
\ee
we get
\bea
&&[Q_{\gamma}, Z^j]=\frac{(-1)^{j}}{32} (\Gamma^jQ)_{\gamma}\nonumber\\
&&[Z^j,Z^k]=\sum_i \frac{(-1)^{i+j+k}}{32}C_i^{jk}Z^i\nonumber\\
&&\{Q_{\alpha}, Q_{\beta}\}=\sum_{k=1,2,5}\frac{1}{k!}(\Gamma^k)_{\alpha\beta}
Z_k
\eea
where $[\Gamma^j,\Gamma^k]=\sum_i C_i^{jk}\Gamma^i$, so we notice that 
$Z^i=(-1)^i\Gamma^i/32$ solves the bosonic part of the algebra. Also, one 
can extend the matrix representation of the bosonic generators by adding an
 extra row and column, corresponding to 
\be
(Q_{\gamma})^{\alpha}=\delta_{\gamma}^{\alpha}\;\;{\rm and}\;\; 
(Q_{\gamma})_{\beta}=C_{\gamma\beta}
\ee
Then the $\{Q, Q\}$ commutator will be just the completeness relation. Note 
that one can independently put a minus sign in front of the $\{ Q, Q\}$ 
commutator and still have a consistent extension of the $Sp(32)$ algebra. 
Equivalently, one can add
$(Q_{\gamma})^{\alpha}$ and $-(Q_{\gamma})_{\beta}$ to the matrix 
representation. More generally, one can multiply $(Q_{\gamma})_{\beta}$ by 
any real constant and generate the same constant in front of the 
$\{ Q, Q\}$, which is therefore still a consistent extension. We will see 
however that for our purposes the - sign suffices. We will also see that 
there is a sense in which we can associate that sign with the 12 dimensional 
chirality of the spinor (if one writes the algebra in (10,2) invariant way).
Note here that one may have thought that one could just redefine the Q's
by and i to change the sign, but that would turn Majorana into anti-Majorana
fermions, and that would be bad, since then $Q_1\pm Q_2$ would have no 
Majorana property.

Finally, in order to get the algebra in a physical form, we rescale the 
$P_{A}$ and $Z_5$ generators by $x=\pm M$. Thus
\be
M_{\alpha\beta}=\frac{1}{x} (C\Gamma^{A})_{\alpha\beta}P_{A} 
+\frac{1}{2} (C\Gamma^{AB})_{\alpha\beta}
M_{AB}+\frac{1}{5!x}(C\Gamma^5)_{\alpha\beta}Z_5
\ee
One can use then the possible 
minus sign and rescaling
in the Q,Q commutator to put back $P_{\mu}$ with plus sign, as 
it is usual. Then one gets the algebra
\bea
&&[P_{A}, P_{B}]=M^2M_{AB}\nonumber\\
&&[M_{AB}, P_{C}]=2P_{[A}\delta_{B ]C}\nonumber\\
&&[M_{AB} , M_{CD}]=4M_{[B}^{[C}\delta_{A ]}^{
D ]}\nonumber\\
&&[M_{AB}, Q]=-\frac{1}{2} \Gamma_{AB} Q \nonumber\\
&& [P_{A}, Q] = \frac{1}{2} x \Gamma_{A} Q \nonumber\\
&& [P_{B}, Z^5_{A_1...A_5}]=i\frac{x}{5!}\epsilon
_{B A_1...A_5 C_1...C_5}Z^5_{C_1...C_5}
\nonumber\\
&& [M_{BC}, Z^5_{A_1...A_5}]=-5 \delta_{[A_1}^{[B}Z^{5C]}
_{A_2...A_5]}\nonumber\\
&&[Z^5_{A_1...A_5}, Q]=\frac{x}{2}\Gamma_{A_1...A_5}Q\nonumber\\
&&\{ Q_{\alpha}, Q_{\beta}\}= (\Gamma^{A} C^{-1})_{\alpha\beta}P_{A}
+\frac{x}{2}(\Gamma^{AB}C^{-1})_{\alpha\beta}M_{AB}+\frac{1}{5!}
(\Gamma^{A_1...A_5}C^{-1})_{\alpha\beta}Z^5_{A_1...A_5}
\label{algebra}
\eea
The general expression for super Poincare extensions, 
(which should therefore contain
also the d'Auria and Fre algebra with spinorial central charge
Q', since it is a good supergroup extension of the Poincare algebra!) is,
for d=11 \cite{vv} 
\bea
&&\{ Q_a, Q_b \} ={\sum _k}' \frac{1}{k!}(\Gamma^k C^{-1})_{ab}  Z^k
\nonumber\\
&&[Q,Z^i]=(-)^i y \Gamma^i Q \nonumber\\
&&[Z^i, Z^j]=2y\sum_k \{ ij;k \} Z^k, Z^{d-k}\equiv \epsilon_d Z^k
\label{compact}
\eea
where $x=\pm M$ (two solutions), i=1,2,5 
and  $\{ ij ; k \}=\frac{i!j!}{s!t!u!}\times$
indices, where $s=(i+j-k)/2, t =(i+k-j)/2, u=(j+k-i)/2$. 

Then $Z_{A}^1=2y/M
P_{A}$ and $Z_{AB}^2=2y M_{AB}$. The rescaling by a real factor 
of the Q,Q commutator can now be seen: just rescale both $Z^i$ and $y$ 
by the same factor $\lambda$, which can be both positive and negative.

The case with spinorial central charge can be obtained as follows.
If one has
\be
[P_{A}, Q]=\Gamma_{A} Q'
\ee
and also 
\be
 [P_{A}, Q']=\Gamma_{A}Q''
\ee
then necessarily \cite{vv} $Q''=\frac{1}{4}x^2 Q$, and so for x=0 
we obtain indeed the algebra we want. Moreover, at x nonzero, defining
$Q_{1,2}=Q/2 \pm Q'/M$ one gets 
\be
[P_{A}, Q_1]=\frac{1}{2} M\Gamma_{A} Q_1, 
[P_{A}, Q_2]=-\frac{1}{2} M\Gamma_{A} Q_2
\ee
which means that $Q_1$ and $Q_2$ are in two independent $OSp(1|32)$ 
groups with different sign ($Q_1$ maps to $Q_1$ and $Q_2$ to $Q_2$). 
Note that if we had the same sign for $Q_1$ and $Q_2$, we couldn't
have extracted linear combinations Q and Q' such that Q maps into 
Q'. So the case of M theory algebra with spinorial central charge 
could appear from two $OSp(1|32)$s with different sign. One still 
needs to check all the algebra.

Let us then describe in detail the contraction of $OSp(1|32)\times 
OSp(1|32)$ to the algebra 
with spinorial central charge. The two groups will have generators 
with indices i=1,2. We saw that $Q_{1,2}=Q/2 \pm Q'/M$
turns two groups with different x's (signs) into the spinor central
charge extension that we wanted. We noted though that minus sign was, up 
to a rescaling of the generators, only nontrivial in a sign in front 
of the Q,Q commutator, so that's what we will keep. We also notice
that  we must now define sums and differences for the other generators
as well. 
Let us then start with $P_{A}=Z_{A}^1-Z_{A}^2$ 
(the x is the same for the two $OSp(1|32)$ groups). 

Since we want to identify
\be
[Z_2^1+Z_2^2, Z_2^1-Z_2^2]=Z_2^1-Z_2^2
\ee
 with $[Z_2, M_2]=Z_2$ (formally expressed, where 
$M_2$ is the Lorentz generator), one can identify
\be
M_2=Z_2^1+Z_2^2, \;\;\; Z_2=Z_2^1-Z_2^2
\ee
Since we also want to identify
\be
[Q_1+Q_2, Z_{2/5}]\sim \Gamma_{2/5}(Q_1-Q_2)
\ee
with $[Q, Z_{2/5}]\sim \Gamma_{2/5}Q'$, it is clear that $Z_2$ obeys
it, and then also $Z_5=Z_5^1-Z_5^2$, and consequently
$Z_5'=Z_5^1+Z_5^2$. 

Then the commutation relations of M are obviously satisfied, and the
only nontrivial thing to check is the $\{ Q, Q\}$ commutation
relation. We want it to be equal to
$Z_1^1-Z_1^2+Z_2^1-Z_2^2+Z_5^1-Z_5^2=P+Z_2+Z_5$, but at  first sight 
we obtain the other set, 
$Z_1^1+Z_1^2+Z_2^1+Z_2^2+Z_5^1+Z_5^2=Z_1+M_2+Z_5'$. But here we remember
that we can have a sign difference in front of the  
$\{ Q,Q\}$ commutator, and that does it.

Before rescaling by M and contracting the group, let's come back to Horava 
and notice that now the group, written in terms of $Q, Q', P, Z_{10}=
\epsilon Z_1, M_2, Z_9=\epsilon Z_2, Z_5, Z_6=\epsilon Z_5'$ is
invariant under parity ${\cal I}$. Its action is to change the sign if
one of the indices in $P, M_2, Z_5, Z_6, Z_9, Z_{10}$ is a 1, and acts 
on Q by $Q\rightarrow \Gamma_1 Q$ and on Q' by $Q'\rightarrow
-\Gamma_1 Q'$. The group commutation relations are invariant under
this transformation.

The reason why it was not invariant beforehand was that $[P_{A}, 
Z_5], [P_{A}, Q]$ and $[Z_5, Q]$ of $OSp(1|32)$ aquire an extra - sign
under parity, $[P_{A}, Z_5]$ because of the $\epsilon$ symbol on its 
r.h.s. (which doesn't change sign), and the Q relations because of the 
$\Gamma_1$. Now the solution is that the Q relations have a Q' on the 
r.h.s., which has an extra minus sign in the transformation, whereas the 
$[P_{A}, Z_5]$ has a $Z_6$, which does change sign, on its
r.h.s. All the other relations continue to be satisfied. 

Finnally, the rescalings of the generators are according to their mass 
dimensions: 

\be
P_A=M^{-1} \tilde{P}_A,\;\; M_2=\tilde{M}_2, \;\; Q=M^{-1/2} \tilde{Q}, 
\;\; Q'=M^{-3/2} \tilde{Q}', \;\; Z_{5,6,9,10}=M^{-1}\tilde{Z}_{5,6,9,10}
\ee
At the level of the action, this is equivalent to rescaling the fields in the 
opposite way (such that the gauge field is invariant), i.e. if considering 
only the vielbein
\be
A=e^AP_A+... =M\bar{e}^A P_A= \bar{e}^A\tilde{P}_A\rightarrow 
[\tilde{P}_A, \tilde{P}_B]=M^2 J_{AB}
\ee
Under these rescalings and upon making the Wigner-Inonu contraction
$M\rightarrow 0$, it is easy to see that Q' and the Zs become
central charges, and only $Z_9$ and $Z_5$ remain nontrivial in the
algebra. ($P, Z_2$ and $Z_5$ are special since they appear in $\{ Q, Q
\}$ and also change Q into Q', whereas $Z_{10}, Z_9, Z_6$ do neither, 
so are rescaled away). Then if one wants of course, one can rescale Q' 
again by another m and it will also dissappear from the algebra. 

But a curious fact is that the numbers in front of Q' on the r.h.s. of
the Q commutators are not the same as the ones in d'Auria and Fre, and 
certainly there are no two solutions. It is either a mistake
in somebody's calculations or (more likely)
a manifestation of the fact that the d'Auria 
and Fre formulation is not completely group theoretical, they needed to 
introduce a 0-form field for the first order formulation of $F=dA_{3}$.

Let us now note that in the end we have
 $M_2=Z_2^1+Z_2^2, Z_2=Z_2^1-Z_2^2$, so that 
$Z_2^1=M_2+Z_2$, and that $Z_2$ is rescaled away and correspondingly 
\be
\omega_1^{AB}=\omega^{AB}(e) +(M)B^{AB},\;\; \omega_2^{AB}=
\omega^{AB}(e) -(M) B^{AB}
\label{composite}
\ee
where $B^{AB}$ is the extra 2-form in d'Auria and Fre, and so
the 3-form $A=B^{AB}\wedge e^A \wedge e^B+...$.
Then in the action $S(\omega_1)+S(\omega_2)$ 
we should vary $\omega_1^{AB}$ and $\omega_2^{ab}$ 
independently, or equivalently $\omega^{AB}(e)$ and $B^{AB}$ independently.
At the linearized level in $B_{AB}$ 
we will get the same from $S(\omega_1)$ as from 
$S(\omega_1)+S(\omega_2)$, since the only difference between $\omega_1 $ 
and $\omega_2$ is the sign of $B^{AB}$.

Let us now try to deduce the supergroup action. We have seen that the gamma 
matrices satisfy the $OSp(1|32)$ algebra, by $Z^i\sim \Gamma^i$. All one needs 
is a group invariant to define the CS action, by 
\be 
d{\cal L}= F^{M_1}\wedge...\wedge F^{M_6}d_{M_1...M_6}
\ee
It is a natural guess that an appropriate invariant would be the trace 
over the gamma matrix representation. One also notices that the trace 
over 11d gamma matrices definitely has the epsilon term  used by Chamseddine:
\be 
Tr (\gamma_{A_1A_2}...\gamma_{A_9A_{10}}\gamma_{A_{11}})
=\epsilon_{A_1...A_{11}}
\ee
But it also contains other terms which will give torsion terms in the 
action. For instance one can easily calculate that
\be
Tr( R^{A_1A_2}\gamma_{A_1A_2}\wedge R^{A_3A_4}\gamma_{A_3A_4}
\gamma_{A_5}...\gamma_{A_8}\wedge T^{A_5}\wedge...T^{A_8})
=2^6 R^{AB}\wedge R^{AB}\wedge (T^{C}\wedge T^{C})^{\wedge 2}
\ee
In a 12 dimensional representation for the operators, one can represent the 
group by gamma matrices as well (more on that in the next section). For the 
AdS subgroup, one could define the action as a trace over 12d spinors, as 
\be
dL_{11}=Tr[(R^{\Pi\Sigma}\Gamma_{\Pi\Sigma})^6]
\ee
but then one wouldn't have the epsilon term ($L_G$), only terms contracted with
delta functions. The result was
called $L_T$ in \cite{btz,tz1,tz2} and $dL_{T, 4k-1}$ contains the 
Pontryagin form in d=2n=4k dimensions, $P=Tr( ({R^a}_b)^n)$ (the n-th Chern 
character), as well as products of lower Chern characters, 
\be
P^{r_1...r_s}=Tr(({R^a}_b)^{r_1})...Tr(({R^a}_b)^{r_s}, \;\; \sum_1^s r_i=4k
\ee
The other choice is to use the trace over only one chirality, which 
reproduces the result for 11 dimensional trace, namely 
\bea 
&&dL_{11}=Tr[(R^{\Pi\Sigma}\Gamma_{\Pi\Sigma})^6(\frac{1\pm \Gamma_{13}}{2})]
=\frac{1}{2} Tr[(R^{\Pi\Sigma}\Gamma_{\Pi\Sigma})^6]
\pm \frac{1}{2}2^6 \epsilon_{\Pi_1...\Pi_{12}}R^{\Pi_1\Pi_2}...
R^{\Pi_{11}\Pi_{12}}\nonumber\\
&&= \frac{1}{2}dL_{T, 11}\pm \frac{1}{2}2^6 L_{G,11}
\eea
But finally, the action which makes most sense is therefore just 
the epsilon term, being the minimal extension of the AdS case of 
Chamseddine. It can be written as 
\be
dL_{11}=Tr[(R^{\Pi\Sigma}\Gamma_{\Pi\Sigma})^6 \Gamma_{13}]
\ee
Up to now we only talked about one of the $OSp(1|32)$ factors. 
Let's propose therefore the bosonic action
\be
dL_{11}=Tr[(R^{(i)}\Gamma^{(i)})^6\Gamma_{13}](1)+ Tr[(R^{(i)}
\Gamma^{(i)})^6\Gamma_{13}](2)
\ee

\section{Type IIB sugra and ``F theory''}

The idea of having an actual (10,2) theory encompassing M theory and type 
IIB, in the spirit of F theory \cite{vafa} was explored before, in 
\cite{ns,tse,hvwp,bdm,bars,hulltime}. 

The (10,2) super Lorentz algebra is $OSp(1|32)$ (see e.g. \cite{dflv}).
For completeness I reproduce here the 12d formulation of $OSp(1|32)\times
OSp(1|32)$. Note that $OSp(1|32)$ can accomodate both IIA and IIB algebras 
as contractions, under certain circumstances \cite{bv}. 
 $OSp(1|32)$ is also the  11d super-Anti-de Sitter group \cite{vv}, since the
generators $P_A, J_{AB}$ of Anti de Sitter have 11 and 55 components
respectively, and $Z^5$ has 462, and the spinor Q has 32. In 12 d,
from the group $OSp(1|32)\times OSp(1|32)$, 
$P_A$ and $J_{AB}$ organize themselves into the 66 components of the 
Lorentz generator $M_{\Pi\Sigma}$, and the extra copy (or rather $Z_9$ and
$Z_{10}$) into a 66 component $Z^{10}$, and the 2 Qs into a 64
component Q, whereas the $Z_5$ and the extra copy (or rather $Z_6$), 
into a 12 d $Z_6$, but that can split into a self-dual part and an
anti-self-dual part. Finally, the 11d Weyl spinors become MW spinors in 
(10,2). 

Let us now describe the $OSp(1|32)\times OSp(1|32)$ 
algebra in 12d (i.e. (10,2)) language; one $OSp(1|32)$ factor suffices. 
The first three relations in (\ref{algebra}) are obviously only 
the dimensional reduction of the $[M_{\Pi\Sigma}, M_{\Omega\Psi}]=
4 M_{[\Pi}^{[\Omega}\delta 
_{\Sigma]}^{\Psi]}$ commutation relation. Since the gamma matrices split as 
$\Gamma_{\Pi\Sigma}=(\Gamma_{AB}, \Gamma
_{A12} )$, where $\Gamma_A=\gamma_A \otimes \sigma_1,
\Gamma_{12}=1\otimes i\sigma_2, \Gamma_{13}=1\otimes \sigma_3$
(notice $(\Gamma_{12})^2=-1$), 
one has that $\Gamma_{AB}=\gamma_{AB}\otimes 1, 
\Gamma_{A 12}=\gamma_{A}\otimes (-\sigma_3)$, and so the next 
two relations in (\ref{algebra}) are satisfied (here $\sigma _3 Q
=\pm Q$). The last two relations come from $[M_{\Pi\Sigma},Z_6]=\delta Z_6$, 
if we remember that because of (anti) selfduality, 
$Z^6_{A_1...A_5 12}=Z^5_{A_1...A_5}$, but $Z^6_{A_1...A_6
}=\pm \epsilon_{A_1...A_{11}}Z^5_{A_7...A_{11}}$. By imposing that 
both the 12d and the 11d charge conjugation matrices act properly 
on gamma matrices, one can find that $C^{(12)}=C^{(11)}\otimes 1$, so 
one doesn't need to specify which dimension C is in. Finally then 
\bea
&&\{ Q_{\alpha}, Q_{\beta} \}=(C\Gamma_2)_{\alpha\beta}Z_2 
+(C\Gamma_6)_{\alpha\beta}Z_6^+\nonumber\\&&
= (C\Gamma_{A12})Z_{A12}+(C\Gamma_{AB})
Z_{AB}+(C\Gamma_{(6)})Z_{(6)}+(C\Gamma_{(5)})\Gamma_{12} Z_{(5)12}
\eea
becomes under dimensional reduction 
\be
\{ Q_{\alpha}, Q_{\beta} \}= (C\gamma_A)(-\sigma_3) P^A
+(C\gamma_{AB})M^{AB} +(C\gamma_{(5)})(\pm 1-\sigma_3)Z_{(5)}
\ee
So we see that the sign of $x=\pm M$ corresponds exactly to the sign of 
$\sigma_3$, i.e. to the 12d chirality. That sign could be rescaled away, 
but as noted before, one could keep it and instead change the sign of the 
Q,Q commutator. 

Let's try to dimensionally reduce to 10d IIB theory. 
It is obvious that one can't have 2 10d spinors of the same 10d
chirality from only one $OSp(1|32)$ factor, 
by definition it has only a 11d spinor,
which splits into two 10d spinors of different chiralities. So 
one needs something with two 11d spinors, and $OSp(1|32)\times 
OSp(1|32)$ does it. This would be a further justification for doubling 
the group, but we will actually 
find that one still needs to break 10d Lorentz invariance, and so in hindsight 
a single 10d spinor would have sufficed. 

One also needs to have something which gives a 10d gravitational theory.
10d IIB seems NOT to be 12 dimensional, but
just to have a (10,2) covariant formulation, the same way as M theory 
seems to be 11d, but have a (10,2) formulation. 

In order to have a consistent truncation of the CS theory to IIB, one must
dimensionally  reduce the equations of motion, not the action. 

Following the general procedure for dimensional reduction explained before, 
one can choose the gauge 
$e^a_{11}=0$, then the field $e^{11}_{\mu} dx^{\mu}$ is a gauge field, 
which can therefore be put to zero as well, and the component
$e^{11}_{11}$ is a scalar field which can be put to a constant. 
The equation of motion (momentarily jumping ahead to use (\ref{bkgr}))
\be
\hat{F}^5 = \tau^5 M^{10} e^{a_1}\wedge ... \wedge e^{a_{10}}
\gamma_{a_1...a_{10}}
\ee
has now no 11d indices. Then the curvature is  
\be
\hat{F}=T^a \gamma^a + T^{11}\gamma_{11}+\bar{R}^{ab}\gamma^{ab}
+\bar{R}^{a11}\gamma^a \gamma^{11}+...
\ee
and one can have the background solution 
\be
\bar{R}^{ab}=\tau M^{2} e^a \wedge e^b , \bar{R}^{a 11}=0=T^a=T^{11}
\ee

In \cite{tse}, Tseytlin found a 12d embedding for the gravity+dilaton+axion 
system of IIB. Namely, 
\be
ds_{12}^2=ds_{10}^2 -e^{-\phi(x)}dy_1^2 +e^{\phi(x)}(dy_2+a(x)dy_2)^2
\ee
implies
\be
\int d^{12}x det E^{(12)}R^{(12)}=\int d^{10}x det e^{(10)} [ R^{(10)}-\frac{1
}{2} (\partial \phi)^2 -\frac{1}{2} e^{2\phi} (\partial a)^2]
\ee
and then the IIB (euclidean) instanton is lifted to a (11,1) gravitational 
wave.

It is natural to ask whether one can do the same now. 
A comment is in order here. The CS theory can of course be written as a 
12d topological theory, with the 11d space as a boundary. But one would
like a usual type of dimensional reduction, at least for the gravity part.

But in our case, 
there is no 12d metric, since there is no 12d vielbein in the
theory. Indeed, the 11d vielbein and spin connection form a ``12d 
spin connection'': $\Omega^{AB}=\omega^{AB}, \Omega^{A12}=e^A$. One
could formally introduce a 12d vielbein $E^{\Pi}$ such that the spin
connection is $\Omega^{\Pi\Sigma}(E^{\Omega})$, i.e. such that one has 
(writing separately the a and 12 components)
\bea
&& dE^A+\Omega^{AB}\wedge E^B +\Omega^{A12}\wedge E^{12}=0
\rightarrow dE^A +\omega^{AB}\wedge E^B +e^A \wedge E^{12}=0
\nonumber\\
&& dE^{12}+\Omega^{12B }\wedge E^B=0 \rightarrow 
dE^{12}+e^B\wedge E^B=0
\eea
In 11d one has (where now everything is a 11d form)
\be
de^A +\omega^{AB}(e^C)\wedge e^B=0
\ee
Then for the 11d forms one can have as solution
$E^A=e^A=\Omega^{A12}$ and  one would need to have $dE^{12}=0$,
while $E^{12}$ would be determined by (now $\omega^{AB}$ is independent of e,
$\omega^{AB}=\omega^{AB}(e)+k^{AB}$)
\be
T^A\equiv de^A+\omega^{AB}\wedge e^B =-e^A\wedge E^{12}
\label{tors}
\ee
When one imposes the constraint $dE^{12}=0$ on this definition of
$E^{12}$, one gets the consistency condition
\be
R^{AB}(\omega)\wedge e^B = (d\omega^{AB}+\omega^{AC}\wedge \omega
^{CB})\wedge e^B =0
\ee
where 
\be
R^{AB}(\omega)=R^{AB}(\omega(e))+Dk^{AB}+k^{AC}\wedge k^{CB}
\ee
Also note that in components (\ref{tors}) reads
\be
E^{12}_{\nu}= -\frac{2e^{\mu}_A}{10}[\partial_{[\mu}e^A_{\nu ]}
+\omega^{AB}_{[\mu}e^B_{\nu ]}]=-\frac{e_A^{\mu}}{10}k^{AB}_{\mu}e^B_{\nu}
\ee
So when the consistency condition is satisfied, $E^{12}$ is related 
to $k^{AB}$. 

It remains to analyze the 12-th components of forms. One has the equations
\bea
&&\partial_{[\mu}E_{12]}^A+\Omega^{AB}_{[\mu}E_{12]}^B +\Omega^{A 12}_{[\mu}
E_{12]}^{12}=0\nonumber\\
&& \partial_{[\mu}E_{12]}^{12}+\Omega_{[\mu}^{12B}E_{12]}^B=0
\eea
and imposing that there is no dependence on the 12th coordinate $\partial_{12}
=0$ one gets 
\bea
&&\partial_{\mu}E_{12}^A +\omega_{\mu}^{AB}E_{12}^B -\Omega^{AB}_{12}e_{\mu}^B 
+e^A_{\mu} E_{12}^{12} -\Omega^{A12}_{12}E_{\mu}^{12}=0\nonumber\\
&&\partial _{\mu}E_{12}^{12}=e_{\mu}^B (\Omega^{12B}_{12}+E_{12}^B)
\eea
%Finally, using the gauge $E^a_{12}=0$ we get 
%\bea
%&&\Omega^{ab}_{12}e_{\mu}^b 
%+ \Omega^{a12}_{12}E_{\mu}^{12}=e^a_{\mu} E_{12}^{12}\nonumber\\
%&&\partial _{\mu}E_{12}^{12}=e_{\mu}^a \Omega^{12a}_{12}
%\eea
Then one gets
\bea
&& \Omega^{12 A}_{12}=e^{\mu}_A\partial_{\mu}E_{12}^{12}-E_{12}^A
\nonumber\\
&& \Omega^{AB}_{12}=\frac{1}{10} e^{\lambda}_C k_{\lambda}^{C[B} e^{A]\rho}
\partial_{\rho} E_{12}^{12} +D_{\mu} E_{12}^{[A}e^{B]\mu}
\eea
and the equation for $E_{12}^{12}$ and $E_{12}^A$
\be
\delta^{AB}E_{12}^{12} +\frac{1}{10} e_{\lambda}^C k_{\lambda}^{C(B}
e^{A)\rho}\partial_{\rho} E_{12}^{12} +D_{\mu}E_{12}^{(A}e^{B)\mu}
=0
\ee
Notice that one can't put $E^A_{12}=0$ since then one will have an equation 
of the type $\delta^{AB}f +v^{(A}w^{B)}=0$, with no solution.

We have therefore 
dimensionally reduced the 12d ``gravitational'' theory to 11d. Before we 
see how to dimensionally reduce to IIB supergravity, 
let's see the dimensional reduction of the $OSp(1|32)\times OSp(1|32)$ 
algebra to 10d IIA and IIB.

By dimensional reduction from M theory, the 10d IIA algebra is 
\bea
\{ Q_{\alpha}, Q_{\beta} \} &=& (C\Gamma_a)_{\alpha\beta}P^a + 
(C\Gamma_{11})_{\alpha\beta}Z+ (C\Gamma_{11}\Gamma_a)_{\alpha\beta}
Z_a + (C\Gamma_{ab})_{\alpha\beta}Z_{ab}\nonumber\\
&&+ (C\Gamma_{abcd}\Gamma_{11})_{\alpha\beta}Z_{abcd}+
(C\Gamma_{abcde})_{\alpha\beta}Z_{abcde}
\eea
with $10+1+10+45+210 +252$ charges corresponding to momentum, 0-brane, 
F1, 2-brane, 4-brane, NS5. By taking the time components and dualizing
them, one would get charges for a  9-brane, 8-brane, 6-brane
and 5-brane (Z has no time component, so no dual charge). 
The last three are easy to identify as the D8, D6 and KK5
monopole, and the 9-brane is also present. 

The analogous process in the M theory algebra gets besides the M2 and
M5 charges also M9 and KK6, so there is a need for a M9, which corresponds
to a cosmological constant. This is another argument for a AdS-CS sugra, 
which has a cosmological constant from the begining.

The 10d IIB algebra is then 
\bea
\{ Q^i_{\alpha} , Q^j_{\beta} \} &=& \delta ^{ij} ({\cal P}C \Gamma^a
)_{\alpha\beta} P_a+ ({\cal P} C\Gamma^a)_{\alpha \beta} \tilde{Z}^
{ij}_a+ \epsilon ^{ij} ({\cal P }C \Gamma^{abc})_{\alpha\beta}Z_{abc}
\nonumber\\
&&+\delta^{ij} ({\cal P} C\Gamma^{abcde})_{\alpha\beta}(Z^+)_{abcde}
+({\cal P}C\Gamma^{abcde})_{\alpha\beta}(\tilde{Z}^+)^{ij}_{abcde}
\eea
with $1= +2\times 10 + 120 +126 +2 \times 126 $ charges, corresponding
to the momentum, F1/D1, D3+, KK5 and NS5/D5. Their duals correspond to 
9 branes (two types), D7. (the charges of the KK5, NS5, D5 are
self-dual). 

The $OSp(1|32)\times OSp(1|32)$ algebra contains (in 11d language)
\be
\{ Q^{i, \alpha} , Q^{j, \beta} \}= (\Gamma_A C^{-1})^{\alpha\beta}
Z^i_A +(\Gamma_{AB} C^{-1})^{\alpha \beta} Z_{AB}^i+ (\Gamma_{ABCDE}
C^{-1})^{\alpha\beta}Z^i_{ABCDE}
\ee
We have to get a nontrivial $\{ \tilde{Q}^1, \tilde{Q}^2 \}$ 
commutator as well as 
$\{ \tilde{Q}^1, \tilde{Q}^1 \}$ different than $\{ \tilde{Q}^2, 
\tilde{Q}^2 \}$, which can only be obtained if we break 10d Lorentz 
invariance (by $\Gamma_0$)
\be
\tilde{Q}^1=(\frac{1+\Gamma_{11}}{2})(Q^1+Q^2), 
\;\;\; \tilde{Q}^2=(\frac{1+\Gamma_{11}}{2})\Gamma_0 (Q_1-Q_2)
\ee
where $\Gamma_0$ is a gamma matrix, not necessarily in the time
direction, and the projectors are there because the IIB algebra has
two spinors of the same chirality. Taking into account that 
\be
\{ {M^{\alpha}}_{\gamma} Q^{\gamma}, {N^{\beta}}_{\delta} Q^{\delta}
\}= {M^{\alpha}}_{\gamma} \{ Q^{\gamma} , Q^{\delta} \} 
{{N^T}^{\delta}}_{\beta} \;\; {\rm and }\;\;
C^{-1}\Gamma_A^T=-\Gamma_AC^{-1}
\ee
one gets 
\bea
&&\{ \frac{1+\Gamma_{11}}{2} (Q^1+Q^2), (\frac{1+\Gamma_{11}}{2})
(Q^1+Q^2) \}= \frac{1+\Gamma_{11}}{2} [ \Gamma_m (Z^1-Z^2)_m
\nonumber\\&&
+\Gamma_0 (Z_1-Z_2)_0 -\Gamma_m (Z^1-Z^2)_{m11} -\Gamma_0
(Z_1-Z_2)_{011} \nonumber\\
&& +\Gamma_{mnpqr}(Z^1-Z^2)_{mnpqr}
-\Gamma_{0mnpq}(Z^1-Z^2)_{0mnpq}]C^{-1}
\eea
Then 
\bea
&&\{ \frac{1+\Gamma_{11}}{2} \Gamma_0(Q^1-Q^2), (\frac{1+\Gamma_{11}}{2})
\Gamma_0(Q^1-Q^2) \}=\frac{1+\Gamma_{11}}{2} [ \Gamma_m (Z^1-Z^2)_m
\nonumber\\&&
-\Gamma_0 (Z_1-Z_2)_0 +\Gamma_m (Z^1-Z^2)_{m11} -\Gamma_0
(Z_1-Z_2)_{011}\nonumber\\
&&  +\Gamma_{mnpqr}(Z^1-Z^2)_{mnpqr}
-\Gamma_{0mnpq}(Z^1-Z^2)_{0mnpq}]C^{-1}
\eea
and 
\bea
&&\{ \frac{1+\Gamma_{11}}{2} (Q^1+Q^2),
\frac{1+\Gamma_{11}}{2}\Gamma_0 (Q^1-Q^2) \}=
\frac{1+\Gamma_{11}}{2} [ \Gamma_0 (Z^1+Z^2)_{11} 
\nonumber\\
&&-\Gamma_m
(Z^1+Z^2)_{m0}+\Gamma_{mn}\Gamma_0 (Z^1+Z^2)_{mn}\nonumber\\&&
+\Gamma_{mn0} (Z^1+Z^2)_{mn}-\Gamma_{mnp} (Z^1+Z^2)_{mnp 11 0}
-\Gamma_{mnpq0} Z_{mnpq 11} ]C^{-1}
\eea

The symmetric traceless tensor has $Z^{11}=-Z^{22}$ and
$Z^{12}=Z^{21}$, whereas the symmetric singlet has $Z^{11}=Z^{22}$, which
allows us to identify these components as follows
\bea
&&(Z^1-Z^2)_m=P_m, -(Z^1-Z^2)_{011}=P_0\nonumber\\
&& -(Z^1-Z^2)_{m11}=Z_m^{11}, (Z^1-Z^2)_0=Z_0^{11}\nonumber\\
&&Z_{mnpqr}=Z^+_{mnpqr}\nonumber\\
&&Z_{0mnpq}={Z^+}^{11}_{0mnpq}\nonumber\\
&&-(Z^1+Z^2)_{m0}=Z_m^{12}, (Z^1+Z^2)_{11}=Z_0^{12}\nonumber\\
&&(Z^1+Z^2)_{mn}=Z_{mn0}\nonumber\\
&&-(Z^1+Z^2)_{mnp110}=Z_{mnp}\nonumber\\
&&-Z_{mnpq11}={Z^+}^{12}_{mnpq 0}
\eea
One should also have something like 
\be
(Z^1+Z^2)_{mn}=M_{mn}, (Z^1+Z^2)_{m0} {\rm or}(Z^1-Z^2)_{m0}=M_{m0}
\ee
Note that one needed to break Lorentz invariance 
in order to reproduce the IIB algebra. One might wonder whether one
hasn't just reproduced T duality. T duality is actually reproduced
(see \cite{bv}) 
if we only have the M theory $\tilde{Q}=Q^1+Q^2$, in which case 
one would say $\tilde{Q}^2=1/2(1+\Gamma_{11})\Gamma_0(Q^1+Q^2)$, and the
only thing which would be changed would be that one would get $Z^1-Z^2
\rightarrow Z^1+Z^2$ in the 1,2 commutator. 
Since $Q^1 $ and $Q^2$ together form a 12d spinor, 
this freedom in T duality should be the 
freedom to reduce on a spacelike or timelike direction first.

Notice that if the susy algebra contracts to the IIB algebra, it is very 
likely that the effective action does as well, and then we know that 
the IIB theory
 can be rewritten in a manifestly 12d form (see \cite{tse}). So it means
there is some way of constructing a 12d metric which dimensionally reduces 
on the torus as in \cite{tse} (i.e. det E on the torus is 1). The 12d 
action however is not just the Einstein action, but it's written in terms 
of $R^{AB}(\Omega)$, and the fake metric is given by $\Omega (E)$. 
One could derive the explicit form of the embedding of IIB in the (10,2) 
theory, but it is cumbersome. 

The advantage over usual T duality is that one has now a highly 
nonlinear action 
on which we perform the procedure, whereas before it was a {\em perturbative}, 
order by order {\em string} duality. It still is a powerful 
statement, even if it breaks Lorentz invariance.

Finally, what are the implications of the fact that the theory has a (10,2) 
formulation?  The uncontracted theory is (10,2) Lorentz invariant (not 
generally covariant since one doesn't have a vielbein, only a spin connection),
and it is only the contraction that breaks this invariance down to (10,1), 
or could be to (9,2). There is then a relation between the (10,1) theory and 
the (9,2) theory which involves one of the scales $M$ or $M_{P,11}$,  
probably even accesible experimentally. Such a relation was also found in 
\cite{hulltime}.

Let us notice here that the CS theory is topological-independent of metric-
yet it is defined on a manifold of definite dimensionality. It is indeed an
extra step to define the theory in (10,2) dimensions, since the forms are now
12d, even if gravity is still 11dimensional. Then it would make sense to 
reduce to a fully covariant (9,2) theory by a different contraction. 
The beauty of this is that it is not a dimensional reduction in gravity, so 
we don't get any ghosts in either case, and both theories are good in their 
own right.

The gauge field is 
\be
A=Me^A P_A +\omega^{AB} J_{AB}+... =\Omega^{\Pi\Sigma}J_{\Pi\Sigma}+...
\ee
where $J_{\Pi\Sigma}$ 
are the generators of $SO(10,2)$. One can now reinterpret the 
fields and split them into $P_A$ and $J_{AB}$ for a $SO(9,2)$ theory instead 
of the $SO(10,1)$. The point is that the AdS group for (10,1) is the same 
as the Lorentz for (10,2), and the dS for (9,2). 

To gain some insight, let's look in less dimensions. In 3d, gravity is of CS
type. However, the gauge group is $SO(2,2)\simeq Sl(2,R)\times Sl(2,R)$, 
which however as we see is completely symmetric is space vs. time. 
The generators are 
\be
J^a=\frac{1}{2}\epsilon^{abc}J_{bc}, \;\; P^a
\ee
forming together $J_{BC}=(J_{bc}, J_{a4}=P_a)$ (as $\Omega^{BC}=(\omega^{bc},
\lambda e^a)$). One can redefine however $P'_a=J_{a1}$ and $\Omega^{a1}=
\lambda  e^a$, which however brings back the theory in exactly the same 
form, since in the (2,1) signature of spacetime one just reinterprets 
space and time. Still, it is encouraging. Note also that one should 
be able to interpret the (2,1) theory as (2,2) Lorentz theory!

In (4,1) dimensions (the next odd dimension), SO(4,2) is the AdS group, 
which is also the (4,2) Lorentz group and the (3,2) dS group. Since by 
dimensional reduction the (4,1) theory gives usual Einstein gravity in 
(3,1)d, the latter should be related also to (3,2) CS gravity (although 
possibly only when the extra dimension is taken into account).

\section{Equations of motion }

Let us now study the equations of motion of the CS action. 
I have calculated more precisely the group invariant in the Appendix. 
The goal of this section is 
to prove that in the limit $M\rightarrow 0$, the 11d supergravity 
linearized equations of motion  solve the (highly) nonlinear CS
supergravity equations of motion. 

The scale M is the AdS scale, and therefore in that limit this (AdS) CS 
supergravity has a background which is almost flat. I will leave the 
discussion of the background for the next section. Now let's focus instead 
on solving the CS supergravity equations of motion in the (almost flat) 
background. 

First, in order to get oriented, let us look at the linearized equations 
of motion in the d'Auria-Fre formulation of 11d supergravity. As we noted,
the symmetry algebra of the CS supergravity in the $M\rightarrow 0$ 
limit coincides with the symmetry algebra of the d'Auria and Fre supergroup 
formulation (up to an unmatching constant, which could be attributed 
to the existence of the zero form). Therefore one expects the d'Auria-Fre
equations to be obtained also from the CS sugra equations in the 
$M\rightarrow 0$ limit.

In the CIS formulation, with curvatures
\bea
&&R^{AB}=d\omega^{AB}-\omega^{AC}\wedge \omega^{CB}\nonumber\\
&&T^A= DE^A-\frac{i}{2}\bar{\psi}\wedge \Gamma^A \psi \nonumber\\
&&\rho= {\cal D}\psi\nonumber\\
&&R^{\Box}=dA-\frac{1}{2}\bar{\psi}\wedge \Gamma^{AB} \psi \wedge E^A\wedge
E^B
\eea
the linearized (linearized in everything but the vielbein)
 equations of motion are
\bea
&&T^A=DE^A=0 \;\;\; (\omega^{AB}\;\;eq.\;\;\;-vielbein\;\;constraint)
\label{constraint}\\
&&{\cal D} \psi =0\;\;\;(\psi \;\; eq. \;\;\; -gravitino \;\;eq.) 
\label{gravitino}\\
&&R^{AC}_{BC}=0 \;\;\; (E^A\;\;eq.\;\;
-Einstein \;\;eq.)
\label{einstein}\\
&&F_{A_1...A_4}E^{A_1}\wedge ...\wedge E^{A_4}=dA
\;\;\;(F_{A_1...A_4}\;\;eq.\;\;\;-constraint)\\
&&{\cal D}_B F^{BC_1C_2C_3}=0\;\;\;(A\;\;eq.\;\;\;Maxwell \;\; eq.)
\eea
When going to the supergroup formulation, 
\bea
&&A=B^{AB}\wedge E^A\wedge E^B\nonumber\\
&&(\pm \frac{1}{4!6!}
\epsilon_{A_1...A_{11}}B^{A_1...A_5}\wedge B^{A_6...A_{10}}\wedge 
E^{A_{11}}+(0/1)\bar{\psi}\Gamma^{A}\eta \wedge E^A)+...
\eea
so $B^{A_1...A_5}$ and $\eta$ do not  appear in the linearized 
equations of motion.

Indeed, now there will be the additional equations
\bea
&&\frac{\delta {\cal L}}{\delta B^{AB}}\simeq \frac{\delta {\cal L}}{
\delta A}\wedge E^A\wedge E^B+...\\
&&\frac{\delta {\cal L}}{\delta B^{A_1...A_5}}\simeq
\frac{\delta {\cal L}}{\delta A}\wedge \pm \frac{1}{4!6!}\epsilon_{A_1...
A_{11}}B^{A_6...A_{10}}\wedge E^{A_{11}}+...\\
&&\frac{\delta {\cal L}}{\delta \eta}\simeq \frac{\delta {\cal L}}{\delta A}
\wedge \bar{\psi}\wedge \Gamma^A E^A+...
\eea
but the last two equations are nonlinear, and only the first remains. 

So the fields $B^{A_1...A_5}$ and $\eta$ don't appear in the linearized 
equations and don't impose additional equations either, but one needs the 
0-form $F_{A_1...A_4}$, and the Maxwell equations are
\bea
&&F_{ABCD}E^A\wedge E^B\wedge E^C\wedge E^D ={\cal D}B^{CD}\wedge E^C 
\wedge E^D \rightarrow \\
&&({\cal D}B^{CD}-F_{ABCD}E^A\wedge E^B)\wedge E^C\wedge E^D=0\\
&& {\cal D}_B F^{BC_1C_2C_3}=0
\eea

Ideally (and that is the first thing d'Auria and Fre tried as well, 
before introducing the 0-form) one would like to obtain the Maxwell 
equation as a equation in terms of forms, by having the 3-form 
$A_{(3)}=B^{AB}\wedge E^A \wedge E^B$ dual to the 6-form 
$B_{(6)}=B^{A_1...A_5}\wedge E^{A_1}\wedge...\wedge E^{A_5}$. 

In general, the duality relation can be written in terms of forms
(without defining explicitly the * operation) by 
\be
\epsilon^{a_1...ad}e_{a_{n+1}}\wedge ...\wedge e_{a_d}\wedge 
F_{(n)}= e^{a_1}\wedge ...\wedge e^{a_n}\wedge (*F)_{(d-n)}
\ee
so one would have liked an equation of motion of the type
\be
{\cal D}B^{CD}\wedge E^C\wedge E^D \wedge E^{A_5}\wedge ...
\wedge E^{A_{11}}\epsilon^{A_1...A_{11}}
={\cal D}B^{B_1...B_5}\wedge E^{B_1}\wedge ...\wedge E^{B_5}\wedge
E^{A_1}\wedge ...\wedge E^{A_4}
\ee
but unfortunately one can only get this equation of motion from a 
lagrangian by multiplying with a zero form (Lagrange multiplier)
$H_{A_1...A_4}$. One could get an equation of motion which reproduces
the above if projected onto vielbeins, for instance
\be
{\cal D}B^{CD}\wedge E^C\wedge E^D \wedge E^{A_6}\wedge ...
\wedge E^{A_{11}}\epsilon^{A_1...A_{11}}
=E^{B_1}\wedge ...\wedge E^{B_4}\wedge {\cal D}B^{B_1...B_4[A_5}\wedge 
E^{A_1}\wedge ...\wedge E^{A_4]}
\ee
and the latter can be obtained from a geometric lagrangian by multiplying 
with $B^{A_1...A_5}$ 
\bea
&&{\cal D}B^{CD}\wedge E^C\wedge E^D \wedge E^{A_6}\wedge ...
\wedge E^{A_{11}}\wedge B^{A_1...A_5}\epsilon^{A_1...A_{11}}
\nonumber\\&&
-\frac{1}{2}E^{B_1}\wedge ...\wedge E^{B_4}\wedge 
B^{A_1...A_5}\wedge {\cal D}B^{B_1...B_4A_5}\wedge 
E^{A_1}\wedge ...\wedge E^{A_4}
\eea
but this lagrangian does not contain the correct Maxwell kinetic term.
We will still see that this lagrangian (and equation of motion) appear 
in the context of CS sugra, in the low energy limit expansion. 

Instead, in our limit (high energy) of 
the CS sugra case, one doesn't have a 0-form, but we still have the 
auxiliary $B^{A_1...A_5}$ so we expect that if 
(\ref{constraint},\ref{gravitino},\ref{einstein}) 
are satisfied and ${\cal D}B^{AB}\wedge 
E^A\wedge E^B=*({\cal D}B^{A_1...A_5}\wedge E^{A_1}...\wedge E^{A_5})$, 
then also the CS sugra equations are satisfied 
(in the $M\rightarrow 0$ limit). Let's prove that.

The gauge fields of the two $OSp(1|32)$ CS factors are (derived from 
the relations between the generators)
\be
2A^{1,2}= M(\pm e^A + Z^A) P_A + (\omega^{AB}\pm M B^{AB}) J_{AB}
+M^{1/2} (\psi \pm M \psi ' )Q+ M(\pm Z_{(5)} + Z_{(5)} ' )M_{(5)}
\ee
and we saw that the bosonic parts of the group invariant $d_{M_1...M_6}$ 
(here M are groups of indices, e.g. in the first line $M_1=[A_1A_2],...,
M_6=A_{11}$) are 
\bea
&&< J_{A_1A_2}...J_{A_9A_{10}}P_{A_{11}}>=\epsilon_{A_1...A_{11}}
\nonumber\\
&&<Z_{A_1...A_5}Z_{B_1...B_5}J_{C_1C_2}J_{C_3C_4}J_{C_5C_6}P_{C_7}>
=T_1\nonumber\\
&&<Z_{A_1...A_5}Z_{B_1...B_5}Z_{D_1...D_5}Z_{E_1...E_5}J_{C_1C_2}P_{C_7}>
=T_2\nonumber\\
&&<Z_{A_1...A_5}Z_{B_1...B_5}J_{C_1C_2}P_{C_3}P_{C_4}P_{C_5}>=T_3
\label{invariant}
\eea
and $T_1$ and $T_2$ were calculated in the Appendix. 

The equations of motion of CS sugra are
\be
F^{M_1}\wedge ...\wedge F^{M_5} d_{M_1...M_6}=0\label{cseq}
\ee
and one sees that because of the different M dependence one can organize the 
equations of motion by powers of M and we will keep only the equations 
up to the first nontrivial order in all the fields (according to the fact 
that only in this limit we expect usual 11d sugra).

We want to put $Z^A=Z_{(5)}'=\psi '=0$ as is the case in linearized 11d sugra,
but one has to remember to use their equations of motion as well.

Then 
\be
2A^{1,2}= \pm M e^A P_A +(\omega^{AB}\pm M B^{AB})J_{AB} +M^{1/2} \psi Q 
\pm M Z_{(5)}M_{(5)} \label{gauge}
\ee
but one has to remember to vary over the fields put to zero as well (consistent
truncation). The equations of motion for each $OSp(1|32)$ factor should be 
satisfied independently (or equivalently, sum and differences, as obtained
by varying w.r.t. e.g., $\omega^{AB}$ and $B^{AB}$, instead of $\omega_1^{AB}$
and $\omega_2^{AB}$).  The number of $P_A$s and $M_{(5)}$s in the group 
invariant appears (as one can check)
 modulo 2, so the sign difference for them is irrelelvant, 
it just contributes an overall sign to the equations of motion. 

The only difference is in $J_{AB}$, but the fields multiplying it have 
different M dependence, so one can easly disentangle the equations. 

The curvature of the gauge field (\ref{gauge}) is 
\bea
&&F^{1,2}= 2(dA + A \wedge A)\nonumber\\
&&= (\pm M T^A +M^2 (B^{AB}\wedge e^B) )P_A 
+ (\bar{R}^{AB}\pm M DB^{AB} +M^2 B^{AC}\wedge B^{CB})J_{AB} 
\nonumber\\&&+
(M^{1/2}{\cal D} \psi \pm M^{3/2} \frac{1}{4} B^{AB}\wedge \gamma^{AB}
\psi)Q\nonumber\\&&
+(\pm M F^{A_1...A_5}+M^2 5B^{A_1B}Z^{BA_2...A_5})Z_{A_1...A_5}
\label{curvature}
\eea
where
\bea
&& T^A=De^A =de^A+\omega^{AB}e^B\nonumber\\
&& R^{AB}=d\omega^{AB}+\omega^{AC}\wedge \omega^{CB}\nonumber\\
&& \bar{R}^{AB}=R^{AB}+M^2 e^A \wedge e^B\nonumber\\
&& {\cal D}\psi = d\psi +\frac{1}{4} \omega^{AB}\wedge \gamma^{AB}\psi 
\nonumber\\
&& F^{A_1...A_5}=DZ^{A_1...A_5} +\frac{1}{5!}M {\epsilon ^{A_1...A_5}}_{B_1...
B_6}e^{B_1}\wedge Z^{B_2...B_6}
\eea
The curvature (\ref{curvature}) has an $M^0$ term, namely the $R^{AB}$
piece. Then the order $M^0$ equation comes from the invariant
\be
< J_{A_1A_2}...J_{A_9A_{10}}P_{A_{11}}>=\epsilon_{A_1...A_{11}}
\ee
where $P_{A_{11}}$ corresponds to the free index, and is 
\be
\epsilon_{A_1...A_{11}}R^{A_1A_2}\wedge...\wedge R^{A_9A_{10}}=0
\ee
$R^{AB}=0$ is only one solution of this equation, however the 
next bosonic equations up to order $M^4$ contain at least one $R^{AB}$
(the order $M^4$ equations contain exactly one), multiplied by many other 
curvatures which we want to be nonzero. That restricts further the
possible solutions, maybe even constraining $R^{AB}=0$ to be unique. 

If $R^{AB}=0$, the order $M^{1/2}$ fermionic equation is satisfied
automatically, the first nontrivial one being at order $M^{5/2}$, where 
all the curvatures are fermionic. And again there are many nontrivial 
equations, up to order $M^4 M^{1/2}$, where we have 4 nontrivial 
bosonic curvatures and a fermionic one. Since I haven't computed the 
fermionic equations, I can't comment on whether these equations 
admit only ${\cal D}\psi=0$ as a solution, but since the equations are 
written in terms of curvatures, it is a solution (even if it's not 
unique). 

Then, if
\be
R^{AB}=0\;\;\;{\rm and}\;\;\; {\cal D}\psi=0
\ee
the first nontrivial equation (and here we will stop, since it is the 
first nontrivial equation for the Maxwell field of 11d sugra) comes at
order $M^5$. The order $M^5$ equations for $B^{AB}$ and 
$Z^{A_1...A^5}$ contain the torsion $T^a$ linearly (from the epsilon 
term and $T_1, T_2$) and cubic order (from $T_3$), and the $M^5$ equation 
for $e^A$ contains the torsion quadratically (from $T_3$). It is again not 
clear whether this is the unique solution, but $T^a=0$ solves the $B^{AB}$ and 
$Z^{A_1...A_5}$ equations at order $M^5$, and then $T_3$ dissappears from 
all equations. 
 
We have found the vielbein contraint, the gravitino equation and the 
Einstein equation. Note that the equations are linearized in all fields 
but the vielbein (gravity). It is still left to check the Maxwell equation, 
and I argued that one expects to find it by a duality relation between
\bea
&&A_{(3)}= B^{AB} \wedge e^A \wedge e^B \;\;\; {\rm and}
\\&&
A_{(6)}= Z^{A_1...A_5}\wedge e^{A_1}\wedge...\wedge e^{A_5}
\eea
At the linearized level, the duality is 
\be
F_{(4)}= DB^{AB}\wedge e^A\wedge e^B= *(F^{A_1...A_5}\wedge e^{A_1}\wedge
...\wedge e^{A_5})
\ee
The CS curvature now is
\be
F^{1,2}= \pm M(DB^{AB}J_{AB}+ F^{A_1...A_5}M_{A_1...A_5})
\ee
and the only remaining equation is the $e^A$ equation 
\bea
&&\epsilon_{A_1...A_{10}C_7}DB^{A_1A_2}\wedge ... \wedge DB^{A_9A_{10}}
\nonumber\\&&
+ 10{T_1}_{A_1...A_5;B_1...B_5;C_1C_2;C_3C_4;C_5C_6;C_7}
F^{A_1...A_5}\wedge F^{B_1...B_5}\wedge DB^{C_1C_2}\wedge DB^{C_3C_4}
\wedge DB^{C_5C_6}\nonumber\\&&
+5{T_2}_{A_1...A_5;B_1...B_5;D_1...D_5;E_1...E_5;C_1C_2;C_7}
\nonumber\\&&
F^{A_1...A_5}\wedge F^{B_1...B_5}\wedge F^{D_1...D_5}\wedge F^{E_1...
E_5}\wedge DB^{C_1C_2}=0
\eea

Substituting the form of $T_1$ and $T_2$, and grouping together 
the epsilon term with the first terms in $T_1$ and $T_2$, then the 
second terms in $T_2$ and $T_3$ and the last terms in $T_2$ and $T_3$,
one gets 
\bea
&&5DB^{C_1C_2}\wedge \{ 
\epsilon _{A_1A_2B_1B_2C_1...C_7} [ DB^{C_3C_4}\wedge DB^{C_5C_6}\wedge
(DB^{A_1A_2}\wedge DB^{B_1B_2}\nonumber\\&&
-2\cdot 12\cdot 50 F^{FGH A_1A_2}\wedge F^{FGH B_1
B_2})\nonumber\\&&
+(50 \cdot 12)^2 F^{IJK C_3C_4}\wedge F^{IJKC_5C_6}\wedge
F^{FGH A_1A_2}\wedge F^{FGH B_1B_2}]
\nonumber\\&&
-500 \epsilon_{A_2...A_5B_2...B_5 C_1C_2 C_7}F^{A_1...A_5}\wedge 
F^{A_1B_2...B_5}(DB^{FG}\wedge DB^{FG}-60 F^{FGHIJ}\wedge F^{FGHIJ})
\nonumber\\&&
-50 \cdot 60 \epsilon_{A_2...A_5B_2...B_5 C_3C_4 C_7}F^{A_1...A_5}\wedge 
F^{A_1B_2...B_5}(DB^{[C_1C_2}\wedge DB^{C_3C_4]}\nonumber\\&&
- 50\cdot 12 F^{FGH [C_1C_2}
\wedge F^{FGHC_3C_4]}\times 3[C_1C_2C_7]) \}=0
\label{maxcs}
\eea
Now let's see how can we satisfy it. If 
\be
dA_{(3)}=*(dA_{(6)})\rightarrow DB^{AB}\wedge e^A \wedge e^B=a*(
F^{A_1...A_5}\wedge e^{A_1}\wedge...\wedge e^{A_5})
\ee
then 
\bea
&&(DK)_{\mu\nu}^{AB}= a{e^{-1}}^{\rho A} {e^{-1}}^{\lambda B} 
{\epsilon_{\mu\nu\rho \lambda }}^{\lambda_1...\lambda_7}F_{\lambda_1
...\lambda_7}\rightarrow \nonumber\\
&& (DK)_{[\mu\nu}^{AB}DK_{\mu ' \nu ' ]}^{AB}=2\cdot 9!\cdot a^2
\delta_{\mu\nu \lambda_1...\lambda_7}^{\mu '\nu '\lambda_1'...\lambda_7'}
F_{\lambda_1...\lambda_7}F_{\lambda_1'...\lambda_7'}\times [\mu\nu \mu '\nu']
7\cdot 12 \cdot 7! \cdot a^2\nonumber\\&&
= F_{[\mu\nu }^{\lambda_3...\lambda_7}F_{\mu'\nu']}^{\lambda_3'...\lambda_7'}
\eea
So that 
\be
DK^{AB}\wedge DK^{AB}= 7\cdot 12\cdot 7! a^2 F^{A_1...A_5}\wedge F^{A_1...A_5}
\label{one}
\ee
Similarly
\bea
&&(dA_3)_{[\mu\nu\rho\sigma} (dA_3)_{\mu'\nu'\rho'\sigma ']}=a^2 
{\epsilon_{\mu\nu\rho \lambda }}^{\lambda_1...\lambda_7}
{\epsilon_{\mu'\nu'\rho' \lambda' }}^{\lambda_1'...\lambda_7'}
F_{\lambda_1...\lambda_7}F_{\lambda'_1...\lambda'_7}\times
[\mu\nu\rho\lambda\mu'\nu'\rho'\lambda']\nonumber\\&&
= 7\cdot 120\cdot 7!\cdot a^2
F^{ABFGH}_{[\mu\nu}F^{CDFGH}_{\mu'\nu'}e^A_{\rho}e^B_{\lambda}e^C_{\rho'}
e^D_{\lambda']}
\eea
so that 
\be
DK^{AB}\wedge DK^{CD}\wedge e^A\wedge e^B \wedge e^C \wedge e^D
=7\cdot 120 \cdot 7! \cdot a^2
F^{ABFGH}\wedge F^{CDFGH}\wedge e^A\wedge e^B\wedge e^C \wedge e^D
\label{two}
\ee
The claim is now that (\ref{one}) and (\ref{two}) imply (\ref{maxcs}). 
It is satisfied if $7\cdot 120 \cdot 7! \cdot a^2= 5\cdot 120\rightarrow
a=1/(7\cdot 12)$. 

Notice though that one doesn't have the totally antisymmetric part of the 
product in the equations of motion, one only has 
$DK^{[AB}\wedge DK^{CD]}$ and $F^{FGH [AB}\wedge F^{CD]FGH}$, so it seems 
like one has to impose some constraints on the fields. 
Note here that one already assumes that only the totally 
antisymmetric parts of $DK^{AB}$ and $F^{A_1...A_5}$ are nonzero, namely
$DK^{AB}\wedge e^A\wedge e^B$ and $F^{A_1...A_5}\wedge e^{A_1}...\wedge 
e^{A_5}$. Indeed, one can easily find that
\bea
&& {DK_{[\mu\nu}}^{[AB}{DK_{\mu'\nu']}}^{CD]}= a_1 {F_{[\mu\nu}}^{FGH[AB}
{F_{\mu'\nu']}}^{CD]FGH}\nonumber\\&&
+a_2{F_{[\mu}}^{FGHI [AB}{F_{\mu'\nu'}}^{D|FGHI|}e_{\nu]}^{C]}
\nonumber\\&&
+a_3(2{F_{[\mu\nu}}^{E_1...E_5}F^{E_1...E_5[CD}e_{\mu'}^Ae_{\nu']}^{B]}
+{F_{[\mu}}^{E_1...E_5[B}{F_{\nu'}}^{|E_1...E_5|D}e_{\nu}^Ce_{\mu']}^{A]})
\nonumber\\&&
+a_4{F_{[\mu}}^{E_1...E_6}F^{E_1...E_6 [D}e_{\nu}^C e_{\mu'}^A e_{\nu']}^{B]}
+a_5F^2 e_{[\mu}^{[A}e_{\nu}^Be_{\mu'}^Ce_{\nu']}^{D]}
\eea
where $a_1=7\cdot 120\cdot 7! \cdot a^2$.
The first line is what one wants, but the next lines have to be constrained to 
be zero. Then we also will constrain DK; one can easily see that one also has
\bea
&&DK_{ABCD}DK_{ABCD}=0, 
{DK_{[\mu}}^{E_1...E_3}DK^{E_1...E_3 [D}e_{\nu}^C e_{\mu'}^A e_{\nu']}^{B]}
=0\nonumber\\&&
(2{DK_{[\mu\nu}}^{E_1E_2}DK^{E_1E_2[CD}e_{\mu'}^Ae_{\nu']}^{B]}
+{DK_{[\mu}}^{E_1E_2[B}{DK_{\nu'}}^{|E_1E_2|D}e_{\nu}^Ce_{\mu']}^{A]})
=0\nonumber\\&&
{DK_{[\mu}}^{E [AB}{DK_{\mu'\nu'}}^{D|E|}e_{\nu]}^{C]}=0
\eea
An observation is that $F^2$ terms ($DK_{ABCD}DK^{ABCD}$ and $F^{A_1...A_7}
F_{A_1...A_7}$) appear on the r.h.s. of the Einstein equation, together 
with $(F^2)_{\mu\nu}$ terms (${DK_{\mu}}^{BCD}{DK_{\nu}}^{BCD}$ and 
${F_{\mu}}^{A_1...A_6}$
${F_{\nu}}^{A_1...A_6}$). We are looking at the linearized
equations of motion, but the Einstein equation has a different M dependence, 
so one expects terms of the type
\be
{DK_{[\mu}}^{BCD}{DK^{BCD}}^{[a}e_{\nu]}^{b]}+(DK)^2 e_{[\mu}^Ae_{\nu]}^B
\ee
to appear in the equations of motion at this order in M.

The next observation is that 
\be
(2{DK_{[\mu\nu}}^{E_1E_2}DK^{E_1E_2[CD}e_{\mu'}^Ae_{\nu']}^{B]}
+{DK_{[\mu}}^{E_1E_2[B}{DK_{\nu'}}^{|E_1E_2|D}e_{\nu}^Ce_{\mu']}^{A]})
\ee
is a symmetric piece, the antisymmetric part appears in (\ref{one})
and can still be nontrivial. The last constraint also leaves the 
possibility of (\ref{one}) and (\ref{two}) to be nontrivial.

\section{``High energy supergravity''}

Up to now I have been talking about only the CS action. But the CS supergroup 
was necessarily an extension of the AdS supergroup. We started with the 
Poincare algebra, which was then extended to $Sp(32)$, but this incorporates 
$SO(10,2)$ as a subgroup, hence the AdS (not dS) signature. 
But for the real world one wants 
to be able to have a flat background, or maybe 
a de Sitter background of small cosmological constant. One could of course 
take the point of view that in the $M\rightarrow 0$ limit, the background 
is flat (as I said), and that somehow the interactions will generate the 
real cosmological constant.

But Horava noticed that there is something one can add to modify the 
background. In a CS theory, one can add Wilson loop observables. Horava's 
observation was that if one has a very large number N of Wilson loops, one 
can make a sort of mean field approximation. Namely, one assumes 
that one looks at path integrals of the type
\be
\int {\cal D} Ae^{-S} \Pi_i Tr {\cal P} e^{\int_{C_i}A}
\ee
and rewrites the result by adding an extra (current) term in the action, i.e.
\be
S=-\frac{1}{g^2} S_{CS} +\int tr (A \wedge {\cal J})
\ee
where ${\cal J}=j_aT^a \delta (C_i)$. Then one  approximates 
the current by a uniform density current, of the type 
\be
J=\frac{\tau^5}{g^2}
 M^{10} \epsilon_{A_1...A_{11}}P^{A_1}\bar{e}^{A_2}\wedge ...\wedge
\bar{e}^{A_{11}}
\ee
where $\tau $ is an arbitrary constant
and one imposes that the normalization of the real and average current 
matches, that is  
\be
\int _{M_{10}} J^0 = c\int _{M_{10}} {\cal J}^0=cN
\ee
Now the action is (taking into account that the CS coupling is 
quantized: $1/g^2=k $ is an integer)
\be
I_{\tau}= -k S_{CS}  -k \tau^5M^{11}
\int  e^{A_1}\wedge... \wedge e^{A_{11}} \epsilon_{A_1...A_{11}}
\ee
(where we have used that $tr(T^A T^B) \propto \delta ^{AB}$)
 with equations of motion (schematically, only for $M_6=A_{11}$ we have the 
second term)
\be
F^{M_1}\wedge ...\wedge F^{M_5}d_{M_1...M_6}
  -\tau ^5M^{10}e^{A_1}\wedge ...\wedge e^{A_{10}}
\epsilon_{A_1...A_{11}}=0
\label{bkgr}
\ee
One can easily check that this gives the background 
\be
\bar{R}^{AB}=M^2\tau e^A\wedge e^B , \;\;
A^{A_1...A_5}=T^A=0
\label{bgr}
\ee
with cosmological constant 
\be
\Lambda=M^2(-1+\tau)
\ee
We have started with a term which gauge invariant (coming from Wilson loops),
so the cosmological extra term $I_M$ in the mean field approximation should 
be invariant as well. Let us check this. 
A general gauge transformation is $\delta A= 
d \lambda ^M T_M + A^P \lambda ^N {f_{PN}}^M T_M$. Since the only 
commutation relations which have $P^A $ on the r.h.s. are M with P 
and Q with Q, one gets that 
\be
\delta e^A= (d\lambda^A +\omega^{AB} \lambda^B) + e^B \lambda^{BA}
+ \psi^{\alpha} (C\Gamma^A)_{\alpha\beta} \lambda^{\beta}
\ee
is the general transformation. One can easily see that the term $I_M$ is 
invariant under $\lambda^A$ (because one can partially integrate 
the D onto another e) and $\lambda^{AB}$ (because the index B has to
be the same as on another one on the e's). The only one which is not 
invariant is the spinor transformation. One can hope therefore that 
there is a supersymmetric version of the cosmological constant term 
which does reproduce the Wilson line. But even the spinor
transformation is an invariance if one assumes that one is working in
the gauge $\gamma^{\mu}\psi_{\mu}=0$, as can easily be checked. That's 
a bit strange, since that is a gauge for the susy transformation 
(spinor gauge transformation, in this formalism). Indeed,
\be
\delta \psi_{\alpha}= (d\lambda_{\alpha}+\omega^{AB}\gamma_{AB}
\lambda_{\alpha}+e^A\gamma_A\lambda_{\alpha})+\lambda^{AB}
\gamma_{AB}\psi_{\alpha}+\lambda^A\gamma_A\psi_{\alpha}
\ee
and we see that $\gamma^{\mu}\psi_{\mu}=0$ imposes the constraint 
\be
\gamma^{\mu}(\partial_{\mu}\lambda_{\alpha}+\omega^{AB}_{\mu}
\gamma_{AB}\lambda_{\alpha}+e^A_{\mu}\gamma_A\lambda_{\alpha})=0
\ee
on $\lambda^{\alpha}$.  

Let's now see what constraints one can put on $\tau$ and k.
One starts with the true equation of motion, 
integrated over a spatial hypersurface,
\be
\int _{M_{10}}F^{\wedge 5}= g^2 \int _{M_{10}}{\cal J}
\ee
Plugging in the background solution ($F=M^2 \tau e^A\wedge e^B \gamma_{AB}$),
one gets the normalization condition 
\be
M^{10} \tau^5 \int _{M_{10}} e^{\wedge 10}=g^2 N
\label{norm}
\ee

Horava also defines M to be the inverse size of the Universe by 
\be
M^{10}  \int _{M_{10}} e^{\wedge 10}=1
\label{mdef}
\ee
If one is looking at the flat space solution $\tau=1$ that implies $1/g^2=N$, 
meaning the quantized Chern-Simons coupling is just given by the number of 
Wilson loops in the system (by what Horava calls a manifestation of the Mach 
principle). 

I will continue to use $1/g^2=N$ although I will not consider either flat 
space as a background
nor M defined as the inverse scale of the Universe. The point is that 
it doesn't seem nice to have two different integers, $k=1/g^2$ and N in the 
theory, especially since one wants to relate this to M theory. The relation
k=N is the most desirable experimentally, but in the low energy expansion 
of section 10 we will study other relations ($k=N^2$ and k=1). More on that 
later, but for the moment let's turn to the definition of 
the effective supergravity.

We have seen that in the $M\rightarrow 0$ limit the 11d sugra equations of 
motion, linearized in everything but the vielbein, solve the CS equations of 
motion. We have also seen that the background with cosmological constant 
solves the CS equations of motion. And the system is invariant under the 
11d sugra supergroup of d'Auria and Fre, so one expects to get the correct 
nonlinear supergravity. The question that remains though, in order to correctly
define the effective supergravity, is what is the value of $M_P$?
One has a mass scale, M, but one also has a dimensionless quantity, N. 

The question is nontrivial, since usually one expands about the quadratic 
part of the action and then $M_P$ is well defined as the strength of 
gravitatonal interactions, but now one doesn't have 
a quadratic action left, after we take the $M\rightarrow 0$ limit. 
For instance, in 11d one could define
the Planck mass $M_P^{(11)}$ by the strength of gravity interaction.
By writing $g_{\mu\nu}=\delta_{\mu\nu}+M_P^{-9/2}h_{\mu\nu}$, 
the Einstein action becomes (schematically)
\be
S=(M_P^{(11)})^9\int d^{11} x R =\int d^{11}x h \Box h
+(M_P^{(11)})^{-9/2}h^2\Box h
\ee
In our case, the $g_{\mu\nu}=g_{0, \mu\nu} 
+h_{\mu\nu}$ expansion implies
\be
R_{\mu\nu}\simeq M^2 g_{0, \mu\nu}+ (h\Box h)_{\mu\nu}+(h^2\Box h)_{\mu\nu}
+...=(R_0+R_1+R_2+...)_{\mu\nu}
\ee
but that $R_1\gg R_0$ as well as $R_1\gg R_2\gg R_3$... So one can't substitute
$R\simeq R_0$, but instead $R\simeq R_1$, so the ``linearized action'' is 
still
\be
S=MN \int (R_1)^5=MN \int (h\Box h)^5
\ee
 Notice that it is hard to introduce the (even linearized)
matter fields as sources for gravity. 
It is also hard to introduce pointlike sources for gravity. Usually 
they are introduced by 
\be
S=\frac{1}{k_N^2} \int d^{11} x R +\int d^{11} x m \delta^{(10)}(x-x(t))
\label{source}
\ee
from which one deduces that the strength of the gravitational interaction 
is given by the Newton constant $k_N^2$, since the sources couple to 
$k_N^2m$, as an alternative defintion of $M_P^9=1/k_N^2$.

But just from the equations of motion of gravity (with no action and no 
sources), the Planck mass given by k doesn't appear.
That is the situation in our case, since 
one just knows the equations of motion 
for the EH  representation, not the corresponding action. And we would try 
to introduce sources by adding the same kind of term as before, but how to 
relate it to a source in the EH equations of motion? It is not clear.

One possible answer though is that the action (\ref{source}) can be 
rewritten in terms of dimensionless variables
\be
S=\int d^{11}(xM_P) ( \frac{R}{M_P^2}+ \frac{m}{M_P}\frac{\delta ^{(10)}}{
M_P^{10}} (x-x(t)))
\ee
and then the coupling to matter is in terms of $m/M_P$, hence $M_P$ is the 
Planck mass. 

In our case, the problem seems to be that one needs a source $[\delta^{(10)}
(x)]^5$ for the Einstein equations, in order to match the usual Einstein 
theory, whereas we have just the usual $\delta^{(10)}(x)$. But notice that 
having an integral 
\be
\int dx (\delta (x))^5=(\delta (0))^4\int dx \delta (x)
\ee
is the same as having $\int dx \delta (x)$ up to the ill defined 
constant $(\delta(0))^4$, which just signals that we need a better description.
So for lack of a better description, we will just replace $\delta(x)$ 
with $(\delta (x))^5$ in the source action, with the understanding that there 
should be a better way to deal with it. 

Then a similar rescaling can be done to obtain massless coordinates, 
and conclude that the 11 dimensional Planck mass is
\be
M_{P,11}=MN
\ee

\section{Cosmological and observational consequences; the cosmological 
constant problem}

Let us now try to see what are the observational consequences  of the 
CS theory. One needs to define now M and N. For Horava, M and N were 
auxiliary quantities going to zero and infinity respectively. But one may take 
a different point of view. One can consider M as a very low energy scale, such 
that it is smaller than any momentum scale one might consider for usual 
experiments, so it has to be a cosmological scale. Then N can be considered 
also to be a total number of partons (to be defined) in the Universe. 
A question that remains open for now in this interpretation is what happens 
in experiments in the lab: who is then N?

Observationally, one could measure $M_{P,11}$ in a lab and cosmologically 
measure the number of particles within the horizon, $N_{HOR}$, the size of
the horizon, $L_0$, and the cosmological constant $\Lambda$. 

Let us review these numbers from \cite{kt}. The Universe is dominated 
by photons today, and the number of photons and baryons in a 
comoving volume stays constant after recombination. 
The baryon to photon ration is 
\be
\eta =\frac{n_B}{n_{\gamma}}\sim 2.7 \times 10^{-8} (\Omega_B h^2)
\ee
and the number of baryons in the horizon is
\be
N_{B_{HOR}}\sim 10^{79} (\Omega_B/\Omega_0^{3/2}h)(1+z)^{-3/2}
\ee
so that the number of photons in the horizon 
\be 
N_{\gamma -HOR}\sim \frac{10^{87}}{2.7} \Omega_0^{-3/2} h^{-3} (1+z)^{-3/2}
\ee
and one then deduces that today (z=0, $\Omega_0\simeq 1$)
\be
N_{0\gamma -HOR}\sim 3 \times 10^{86}
\ee
which is the number of particles in the horizon today.
For the horizon distance we are going to use the 
Hubble distance $L_0=cH_0^{-1}\sim 
10^{28}cm$ and since 
 $1 GeV^{-1}\sim 2 \times 10^{-14} cm$,  $L_0 \sim .5 \times 10^{42}
GeV^{-1}$. The definition of horizon distance varies a bit: for a
time evolution of the scale factor of
$R(t)\sim t^n$, $H^{-1}=t/n$,  the proper distance to the horizon is 
$d_H(t)\sim t/(1-n)=H^{-1} n/(1-n)$, but since n=1/2 in the radiation 
dominated era (R.D.) and n=2/3 in the matter dominated era (M.D.), the 
errors are negligible at this level of accuracy.
Finally, the measured cosmological constant  $\Lambda\sim 10^{-123}M_{P,4}^4$.

A few elements of FRW cosmology are also needed.  
For a fluid with equation of state $p= w \rho$, 
energy conservation gives 
\be
\rho \sim R^{-3(1+w)}
\ee
If that particular fluid dominates the overall energy, one gets the behaviour
\be
R\sim t^{\frac{2}{3(1+w)}}
\ee
which means that the dominant fluid always behaves like $\rho \sim t^{-2}$.
Moreover, if $\Omega \simeq 1$ and $2/(3(1+w))=n$, then 
\be
\rho \simeq\frac{3}{8\pi G}\frac{n^2}{t^2}\sim \frac{M_P^2}{L_0^2}
\ee
and since today $\rho_0\sim \rho_{\Lambda, 0}$, we have $\rho_{\Lambda , 0}
\sim M_P^2 /L_0^2$.

Experimentally, one knows that the Universe is accelerating, that means that 
in terms of an effective fluid with w, today we have
$w\leq -1/3$. The equality
 corresponds to $\rho \sim R^{-2}, R\sim t$ (the last 
only if it dominates). 

We saw that the background had a small cosmological constant, 
\be
R^{AB}=(1-\tau)M^2 e^A\wedge e^B
\label{backgr}
\ee
so in the generic case ($\tau$ of order one),
$\Lambda \sim M^2$ (or rather, $\Lambda\sim M^2 M_{P,11}^2$ in usual 
notation). Also note that if $\tau >1 $, we have a dS solution, as in 
the real world ($\tau =0$ gives AdS). 
Assuming now that the observed cosmological constant comes 
entirely from 11d , and also that it does not come from some CS quantum 
effect, that is everything, and the compactification to 4d just gives 
the common volume factor changing $M_{P,11}$ to $M_{P,4}$.
Therefore $\Lambda_4\sim M^2 M_{P,4}^2$, and experimentally 
$\Lambda\sim 10^{-123}M_{P,4}^4\sim M_P^2/L_0^2$, so
\be
M\sim 10^{-61.5}M_{P,4}= 10^{-42.5} GeV
\ee
which remarcably is just equal to $1/L_0$ ($L_0=10^{42} GeV^{-1}$)
since $\rho_{\Lambda ,0}\sim M_P^2/L_0^2$, 
and $1/L_0$ is the absolute minimum momentum in the Universe (its wavelength 
can't be larger than the horizon)!
Moreover, the matter coupling (obtained by adding matter as extra Wilson 
lines)
\be
\int_{C_i}A=\sum _{i=1}^k M \int d^{11}x \sqrt{g} \delta ^{(10)} 
(x-x_i(t))
\ee
has coupling constant $\sum_i M=kM$, quantized in units of the minimum
momentum, as it should! So all momenta in the Universe are 
larger than M, meaning that M corrections can be neglected for everything 
but cosmology. 

One sees therefore that if one reverses the logic and imposes that $M=1/L_0$,
one obtains the correct order of magnitude for the cosmological constant!
And the only assumption was that the cosmological constant was already present 
in 11d, so the result is independent of the particular model 
presented here ($OSp(1|32)\times 
OSp(1|32)$ CS sugra). It would be valid for any model with an 11d background
given by (\ref{backgr}).

The next step is to try to relate the partons with the number of particles 
in the Universe. Note that the partons  were defined as Wilson lines, 
and they carry $OSp(1|32)$ indices on the worldline. They also define 
space in the sense of the Mach principle, so it makes sense to speculate
a relation to the BFSS definition \cite{bfss} 
of M theory as Matrix theory of D0 branes \cite{sen,seiberg}. 
In that case the gauge indices were of an auxiliary SU(N) group giving 
the spacetime diffeomorphisms, and in our case the spacetime metric 
is generated by a mean field theory, but the similarity is suggestive of a 
connection. One should note in here that in the absence of the partons (Wilson
lines), there is still a spacetime, of AdS type. But, on one side, we are 
talking about the whole Universe as a system, 
and the cosmological solution is derived 
only once we specify $\Lambda$, hence the partons. And even if we were talking 
about Earth based experiments, 
so that the space is approximately flat and $\Lambda $ is irrelevant, we 
could take the point of view that any source matter can be described by 
Wilson lines, so the spacetime is determined only once we define the matter.

Let's see if we can find out anything without specifying a 
precise relation. The only thing one needs is that the partons can 
be identified with D0 branes. In that case, N is the rank of the Matrix 
theory gauge group, and then there are of the order of $N^2$ string states,
which can be associated with physical particles (mesons, gauge 
fields,...) in the system. So the concrete implication of the identification 
of D0 branes and partons is that $N^2\sim N_{HOR}$

Then 
\be 
M_{P,11}=MN= MN_{HOR}^{1/2}\sim 10^{-42.5}GeV \times 10^{43.5}\sim 10 GeV
\ee
which is a bit low, but at this level of accuracy it could still be identified 
with the popular TeV scale, so presumably accessible to accelerator 
experiments. This in turn would mean that the average inverse radius of the 
compact dimensions would be $R^{-1}\sim 30 MeV$ (from $M_{P,11}^9R^7=
M_{P,4}^2$). 

Let us now note that in a cosmological context neither the definition of M 
as the size of the Universe 
in (\ref{mdef}) nor the normalization condition (\ref{norm}) need hold. 
The size of the (spatial) universe evolves with time, 
so one could only define the size at a time $t_0$ and set it to 1. 
It is natural to set it at an early time.
But why can one have violation of the normalization equation (\ref{norm})?
More precisely, it's not (\ref{norm}) that can be violated, but one can have  
\be
\int_{M_{10}} J=N_{HOR}^{\alpha}
\label{alpha}
\ee
with $\alpha \neq 1$. Really one is just counting the number of Wilson 
lines inside $M_{10}$, so it would seem that would have to be by definition 
just $N_{HOR}$? The answer to that is very simple. Before solving the 
CS equations of motion 
one doesn't have a metric, therefore there is no notion of distance 
or time, and hence no cosmological expansion either. 
The only thing that we have at our disposal is a number of Wilson lines in a 
manifold with no metric. So how does one even define the hypersurface $M_{10}$ 
which is located at a time t? One has to define a priori t somehow, and the 
only way is via a number of Wilson lines. So define t by the fact that 
one has for each physical particle a number of Wilson lines defining the 
volume occupied by it. Since the number of particles inside the horizon at 
time t will also be a function of t, the number of Wilson lines inside 
the horizon will be $N_t= N_{HOR}^{\alpha}$, with $\alpha > 1$ (the most 
natural would be I guess $\alpha =2$, i.e. time is measured by $N_{HOR}$, 
but any definition will do).

If one insists on having $g^2N=1$ as until now, then the two conditions 
above are one and the same. Let us see whether there is a consistent 
cosmology we can derive. The spatial size of the Universe (up to the 
horizon) can be calculated, so at present 
\be
M^{10}\int_{M_{10}} e^{\wedge 10}\sim M^{10} (L_0)^3R^7= (ML_0)^3 (MR)^7=
10^{-287}
\label{cosmo}
\ee
Unfortunately, this means that in (\ref{alpha}), $\alpha<0$, which is hard 
to explain, since it would mean that there is less than one Wilson line 
in the whole Universe, which is clearly wrong. 

Moreover, if one tries to impose $M\sim 1/L_0$ at all times (so that M is 
a dynamical scale for the Universe), then $M\sim 1/t$, and then 
also $\Lambda\sim M^2\sim t^{-2}$, which interestingly enough is just the 
behaviour of the dominant component of the energy density in a FRW cosmology.
One could speculate at this point that the dark energy and dark matter are 
therefore related in this scenario. However, the time evolution of the 
number of particles within the horizon is given in a FRW cosmology by
\be
N\sim t^3/R(t)^3 \sim t^{\frac{1+3w}{1+w}}
\ee
Here I have assumed that the number of particles in a comoving volume
is constant, which is approximately true in FRW cosmology. 
More precisely, both in the matter dominated (M.D). and in the radiation
dominated (R.D.) phases the entropy is 
conserved (and hence the number of photons; the two are related by 
$s=1.80 g_{*s} n_{\gamma}$, where $g_{*s}$ has order 1 jumps at 
phase transitions, but remains of order one).

Then one has 
\be
N_H\sim t (M.D.) \;\;\; N_H\sim t^{3/2} (R.D.)
\ee
and then $M_{P,11}=MN \sim MN_H^{1/2}\sim t^{-1/4}$(R.D.) and $\sim 
t^{-1/2}$ (M.D.). But experimentally, 
$M_{P,11}$ cannot have such a drastic time evolution. 
Even if $M_{P,4}$ is constant, that would mean that the scalar fields 
assocaited with the compact space evolve in time, and there are stringent 
constraints on that. 

Let us now try to fix (\ref{cosmo}). Possible solutions include a higher 
M or a higher internal volume. Let us relax the condition $g^2 N=1$. 
Then $M_{P,11}=Mk$ ($k=1/g^2$), and so 
\be
\int M^{10}\int_{M_{10}} e^{\wedge 10}\sim \frac{10^{164}M[GeV]}{k^9}
\ee
One could increase M, but then loose the interpretation as the smallest 
momentum in the Universe. And if one decreases k then one also decreases 
$M_{P,11}$, which is already at the experimental limit, so that is not good 
either. 

So unfortunately it seems that there is no good cosmology we can have, at 
least not if we insist on (\ref{alpha}) with $\alpha >1$. 

Finally, what is the meaning of M and N in M theory? They can be understood as 
just extra parameters needed to define the whole Universe by a Mach principle 
argument. Usual M theory is understood as the high energy limit of the CS. 
In a usual definition of M theory, one has $1/M_{P,11}$ corrections and 
1/N corrections. Here M and N seem fundamental, but if one takes $M_{P,11}$ as 
fundamental, then M corrections would be $M_{P,11}/N$ corrections. The precise 
relation to M theory is not clear though.

\section{Quantum theory?}

It is not clear how to do a good quantization of the CS theory. 
In his paper on 3d gravity as a CS theory \cite{witten}, Witten pointed 
out that the short distance behaviour is governed by the expansion around 
the trivial vacuum (``unbroken phase'') 
$A=0\rightarrow e=\omega=0$. There is a quadratic action 
around that vacuum, and the theory is renormalizable, therefore the expansion 
around nontrivial vacua makes sense as well. But he also pointed out that 
doens't happen in 4d, since there is no quadratic action around the 
trivial vacuum, since the Einstein term is $\int e e d\omega +...$ That's why 
the theory is nonrenormalizable. The same will be true in 11d. But the 
point is that one doesn't expect CS gravity to be renormalizable, but rather 
CS supergravity, probably susy being the key. One just can't use the expansion 
around the trivial vacuum as an argument, but instead one must use
 the fact that the usual 
expansion gives 11d sugra. One can say though that if the theory is 
renormalizable, there will only be a finite number of terms which can be 
added to the theory. Moreover, if the theory is a gauge theory, one will 
have to preserve the gauge symmetry after quantum corrections. One will 
probably not generate any new terms in the action, although a priori there
could be $\int F^n$ terms, for instance. But these would need the definition 
of an inverse metric and the star operation (nongeometric), so it is likely
they are absent. A possible hole in the argument is the fact that one
puts a matter term in the action, which is of a different form.

If this argument is true, then a prediction of M theory would be the 
CS form of the quantum corrections. A classical background will satisfy the 
CS equations of motion, so the Einstein background could be good, but 
quantum corrections could move us away from this. So the interaction terms 
obtained by expanding around the background will involve both the interaction 
terms in 11d sugra and interaction terms in the M theory 
quantum corrections. The 
latter will be the ones having less powers of $M_P$ (extra powers of 
$\alpha '$). 

Yet a good possibility is that the expansion in M (the expansion parameter 
of the CS) obscures not only the CS origin of the classical 11d sugra, but 
also of its quantum corrections ($1/M_P^2$ expansion). Let's try to see 
which case happens. 

We noted already that the Zwiebach conjecture that LL terms are the only ones 
allowed by string theory is not quite true. 
Also \cite{ggv} showed  that the M theory 
calculations reproduce the $R^4$ term in string theory, which is of the 
form $\int t_8t_8R^4$, where
\bea
&&t_8^{ijklmpq}= 4^{-4}\gamma^{ij}_{a_1a_2}...\gamma_{a_7a_8}^{pq}
\epsilon^{a_1...a_8}=tr_{S_0}(R_0^{ij}R_0^{kl}R_0^{mn}R_0^{pq})
\nonumber\\
&& = -\frac{1}{2}\epsilon^{ijklmnpq}-\frac{1}{2}
((\delta^{ik}\delta^{jl}-\delta^{il}\delta^{jk})(\delta^{mp}
\delta^{nq}-\delta^{mq}\delta^{np})+...)
+\frac{1}{2}(\delta^{jk}\delta^{lm}\delta^{np}\delta^{qi}+....)
\eea
In here though, the tensors are all 8d lightcone tensors.
When talking about covariant 10d terms, one refers to the 
$t_8$ tensor as the above without the epsilon symbol. 

1) The heterotic string SO(32) action has the gravitational terms \cite{tse2} 
\be
S=-\frac{1}{8}\int [ R-\frac{1}{8} tr R^2+b_1 (tr R^2)^2 
+b_2 (t_8 t_8 R^4-\frac{1}{8} \epsilon_{10} \epsilon_{10} R^4)]
\label{heterotic}
\ee
where $trR^2=R^{ab}_{\mu\nu}R^{ba \mu\nu}$ can be modified by a local field 
redefinition to the LL term ${\cal L}^2$ \cite{mt} and the tensor $t_8$ 
is defined by for instance (here the order matters because of the 
SO(32) gauge indices)
\bea
&&t_8 F^4= 16 F^{\mu\nu}F_{\rho \nu}F_{\mu\lambda}F^{\rho\lambda}
+8F^{\mu\nu}F_{\rho\nu}F^{\rho\lambda}F_{\mu\lambda}\nonumber\\
&&-4F^{\mu\nu}F_{\mu\nu}F^{\rho\lambda}F_{\rho\lambda}
-2F^{\mu\nu}F^{\rho\lambda}F_{\mu\nu}F_{\rho\lambda}
\rightarrow
t_8t_8 R^4=24 t_8 [trR^4-\frac{1}{4}(tr R^2)^2]
\eea
The epsilon term (Gauss Bonnet topological 
term in 8d) can't be calculated from the 4 point scattering of 
gravitons since as we saw these terms have no leading term so the 
$R^4$ term would contribute to the 5 graviton amplitude.
It can be fixed by comparison with the sigma 
model beta function. 

2) The IIA string action on the other hand doesn't have $R^2$ and $R^3$ terms 
and reads (only gravitational terms) \cite{tse3}
\be
S=-\frac{1}{2k_{10}^2}\int e^{-2\phi}( R+b_0 \alpha '^3 J_0)
-\frac{1}{2\pi \alpha '} \int b_1 {\cal J}_0
\ee
where $J_0=t_8t_8R^4+1/8 \epsilon_{10}\epsilon_{10}R^4$ and 
${\cal J}_0=t_8t_8R^4 -1/8 \epsilon_{10}\epsilon_{10}R^4$.

The superinvariant which has an ${\cal N}=$IIA extension 
and can be thus embedded into M theory is 
\be
\int b_1(J_0-2{\cal I}_2)
\ee
where ${\cal I}_2 =1/8 \epsilon_{10}\epsilon_{10}R^4$. Then the 11d 
action has the gravitational terms
\be
S=-\frac{1}{2k_{11}^2}\int R-b_1T_2\int (J_0-2{\cal I}_2)
\ee
where the $J_0$ term can be determined from the 4-graviton amplitude 
with M-theory cut-off (one loop calculation in eleven dimensions 
\cite{ggv,rt}), and the ${\cal I}_2$ term is the superpartner of the 
$C_3 trR^4$ term in 11d (calculated from M5 brane anomaly cancellation 
condition). Note that now ${\cal I}_2=1/(4\cdot 3!)\epsilon_{11}\epsilon_{
11} R^4$ and $
{\cal I}_2 e^0\wedge ...\wedge e^{10}=2/3\epsilon_{11}
e^{\wedge 3}\wedge R^{\wedge 4}$

3) The M theory quantum gravitational action is therefore
\be
S=-\frac{1}{(2\pi)^5 l_P^9}\int (R -\frac{l_P^6 }{3^2 2^{15}}
(\frac{1}{3!}\epsilon_{11}\epsilon_{11}R^4+4 t_8t_8 R^4))+...
\ee

By comparison, the CS gravitational action is (schematically, the contraction
of tangent indices is with an epsilon tensor; it is in first order form unlike 
the above)
\be
S=(...)\frac{5}{9}\int [R\wedge e^{9}+\frac{9\cdot 2}{7}
R^{\wedge 2}e^{\wedge 7}+\frac{9\cdot 2}{5} R^{\wedge 3}\wedge 
e^{\wedge 5}+3 (R^{\wedge 4}\wedge e^{\wedge 3})+\frac{9}{5}
R^{\wedge 5 }\wedge e]
\ee

Among these, the $R^4$ LL term (in brackets) is, in metric notation
\be
S=\int \sqrt{g}R_{\mu_1\mu_2}^{\nu_1\nu_2}R_{\mu_3\mu_4}^{\nu_3\nu_4}
R_{\mu_5\mu_6}^{\nu_5\nu_6}R_{\mu_7\mu_8}^{\nu_7\nu_8}
\delta^{[\mu_1...\mu_8]}_{\nu_1...\nu_8}
\ee
A different CS-type term (with a different trace than the epsilon
one, for instance the Pontryagin term $\int_{M_{12}}(R^A_B)^6)$ would
have, in 11d metric notation at least an extra $T^a$ and/or $\omega^{ab}$
and be contracted with (among others) 
\be
\epsilon^{\mu_1..\mu_{11}}e_{\mu_{11}}^{a_{11}}...\sim |e|
\epsilon^{a_1...a_{11}}{e^{-1}}^{\mu_1}_{a_1}...{e^{-1}}^{\mu_{10}}_{a_{10}}
\ee  
So apparently there are more terms in string
theory than in the CS. However, these might come from quantum corrections 
only. The problem is that one has M theory
terms which are not even forms (so not of CS form), namely there 
is no epsilon tensor in their definition, in particular the $t_8t_8R^4$
term. 

Note now that there are no $R^2$ and $R^3$ corrections 
allowed by susy in type II 10d perturbative string theory. 
Indeed, if one expands the tree-level amplitudes 
for 3 and 4 gravitons in type II/I theories, one doesn't see any such terms, 
only in the heterotic model (with less susy) we saw an $R^2$ term 
(\ref{heterotic}). 
Also, it is proven that any loop amplitude with 3 or fewer massless particles
is zero. In particular that means that there is no renormalization of the 
Newton's constant, since that can be measured from the 3-graviton amplitude 
(from the Einstein term). 

But clearly a $R^2$ or $R^3$ CS interaction in 11d would generate a 
corresponding one in 10d, so how do we reconcile this with the statement about
existence of such terms in the CS action? Supersymmetry 
restricts us, so if we reproduce sugra at high energies, then susy should 
again dictate the absence of such terms, at least at high enough energies. 

In conclusion, the assumption that somehow the M expansion obscures not 
only the CS form of the 11d sugra, but also of its quantum corrections, is 
the only remaining option.

\section{Digression: what if CS sugra is a low energy expansion?}

In \cite{banados} 
the implicit assumption was made that one expanded around large M the 
$OSp(1|32)$ CS action. It was claimed that one expanded around small M, but
by linearizing the equations of motion one was in effect expanding around 
large M, since the $OSp(1|32)$ curvature 
\be
F=(R_0^{AB}+R^{AB})\gamma_{AB}+...
\ee
where $R_0^{AB}=M^2 e^A\wedge e^B$. Saying that we take the small M 
limit is the same as having $M^2$ much smaller than any momenta characterizing
the size of $R^{AB}\sim p^2$, so one can't expand around $M^2 e^A \wedge e^B$. 
But the calculations (if not the conclusion) of \cite{banados} 
can still be valid
in a certain regime. Namely, M could be large, but the cosmological constant 
small. 

That is, take the point of view advocated by Horava, namely that one wants 
to have flat space as a solution of the CS sugra, and so $\tau \simeq 1$. 
But the current $J$ proportional to $\tau^5$ was obtained from a mean field 
approximation of a discrete current. So it is natural to assume that in 
a more precise treatment we would get ${\cal O}(1/N)$ corrections to $\tau$. 
So
\be
\Lambda=M^2(1-\tau)\sim \frac{M^2}{N}
\ee
and given that N is very large, $\Lambda$ could still be within the 
experimental constraints, and M be as low as 1TeV experimentally (that is 
where new physics is expected). And since $\Lambda/M_{P,4}^2\sim 10^{-123}$
\be
N\sim \frac{M^2}{\Lambda}\geq 10^{91}
\label{condition}
\ee
which incidentaly is very close (relatively speaking) to $N_{HOR}$.

So it is possible to have a large M and still a very small $\Lambda$. 
Then \cite{banados} have the curvature
\be
F =M T^A \gamma_A +
\bar{R}^{AB} \gamma_{AB}+\frac{1}{5!}
\bar{F}^{A_1...A_5}\gamma_{A_1...A_5}+Q D\bar{\psi}
\ee
The background is $\bar{R}^{AB}=M^2\tau e^A\wedge e^B, A^{A_1...A_5}
=T^A=0$, and since $\tau - 1\ll 1$ (as opposed to $1-\tau \sim 1$),
we can still expand around it as in \cite{banados}. 
The Einstein equation one obtains then  is 
\be
\tau^4M^8\gamma_8\wedge (\delta R^{AB}+2M^2(1-\tau)
e^A \wedge \delta e^B )\gamma_{AB}=0
\ee
which is exactly the linearized equation for the background.
Here we have used that if $\gamma_n=e^{A_1}\wedge
... \wedge e^{A_n} \gamma_{A_1...A_n}$, then $ \gamma_n \wedge
\gamma_m =\gamma_{n+m}$. 

Then $\omega^{AB}=\omega^{AB}(e)+k^{AB}$, which means that 
$T^A=k^{AB}\wedge e^B$ and
\be
\delta R^{AB}=\delta R^{AB}(\omega (e))+Dk^{AB}
\ee
The fermionic  equation is 
\be
\gamma_8\wedge D\psi=0
\ee
and the bosonic equations are
\bea
&& \epsilon_{A_1...A_{11}} e^{A_1}\wedge ...\wedge e^{A_8} \wedge 
\delta R^{A_9 A_{10}}=0 \nonumber\\
&& \epsilon_{A_1...A_{11}} e^{A_1}\wedge ... \wedge e^{A_6}
\wedge (M e^{A_7} \wedge e^{A_8} \wedge T^{A_9}- \bar{F}^{A_7A_8A_9BC}\wedge
e^B\wedge e^C
)=0\nonumber\\&&
e^{[A_1}\wedge ... \wedge e^{A_4}\wedge \bar{F}^{A_5]B_1...B_4}\wedge e^{B_1}
\wedge...e^{B_4}
+\frac{100}{7} \epsilon^{A_1...A_{11}}e_{A_6}\wedge ... \wedge
 e_{A_{11}} dA_{[3]}=0
\label{loweq}
\eea

If $\delta R^{AB}(\omega (e))=0$, the first (Einstein) equation in 
(\ref{loweq}) is 
just $(d * A_{3})\wedge e^{A_{11}}=0$, which is just the Lorentz condition, 
whereas the second and the last (multiplied by $\wedge e^{A_5}$) are 
\bea
&& \frac{7M}{3} A_{[3]}=* d A_{[7]}\nonumber\\
&& dA_{[6]} +M A_{[7]} =50 * d A _{[3]}\nonumber\\
&&A_{[7]}\equiv \frac{1}{6!5!} \epsilon_{a_1... a_{11}}
A^{a_1...a_5}\wedge a^{a_6}\wedge ... \wedge e^{a_{11}}
\eea
which implies (by eliminating 
$A_{[7]}$), that  $*d(*d A_{[3]}-dA_{[6]})=*d*d A_{[3]}=5/150 M^{2} A_{[3]}$,
so unfortunately the 3-form gets a very large mass. But it is remarcable 
that the system still contains the same fields (however now only from an
$OSp(1|32)$ factor), and one still gets the correct Einstein equation. 

So let us explore for a moment the possibility that 11d sugra is the low
energy expansion of a CS sugra (not necessarily of the $OSp(1|32)$ type,
but with the same gravitational action)
and see what cosmological implications does it have. 
First of all, from (\ref{condition}) one can see that one can't have $N=N_{HOR}
^{1/2}$ as before, so the simplest assumption is that the partons would have 
to be physical particles, so $N\sim N_{HOR}$. 

Then the Planck scale is now defined in the usual way, as the coefficient 
of the (dominant) R term in the action
\be
S=\frac{M^9}{g^2}\int d^{11}x \; det (e)(\frac{M^2}{N} +R+{\cal O}(R^2))
\ee
Therefore 
\be
M_{P,11}=\frac{M}{g^{2/9}}=MN^{1/9}
\ee
and if 
\be
M=M_P\sqrt{N\frac{\Lambda}{M_P^2}}\sim 10^{19}GeV\sqrt {10^{87-123}}\sim 10 GeV
\Rightarrow M_{P,11}\sim 10^{7.5}TeV 
\ee
Then the average size of the compact directions is
\be
R^{-1}\sim M_{P,11}(\frac{M_{P,11}}{M_{P,4}})^{2/7}\sim 10^4 TeV
\ee

During R.D. the comoving scale
 $R(t)\propto t^{1/2}$, so the number of particles 
$N=N_{HOR} \propto t^3/R(t)^3\propto t^{3/2}$, so 
\be
\int_{M_{10}} e^{\wedge 10}\propto N^2 \;\;\; (R.D.)
\ee
and during M.D. $R(t)\propto t^{2/3}$, so $N\propto t^3/R(t)^3 \propto t$, 
\be
\int_{M_{10}} e^{\wedge 10}\propto N^3 \;\;\; (M.D.)
\ee
But the Universe spends much more time in the R.D. phase: in the M.D. 
phase there is only a size change of 
\be
z_{eq}\simeq \frac{R_0}{R_{eq}}\simeq 10^4
\ee
meaning a change in the number of particles in the horizon of 
($N\sim t \sim R^{3/2}$)
\be
\frac{N_0}{N_{eq}}\simeq (1+z)^{-3/2}\simeq 10^6
\ee
I mentioned that the value of $M^{10}\int e^{\wedge 10}$can be fixed 
at an early time, after which it evolves. 
Therefore one can take the R.D. proportionality result and define it 
as an equality in the cosmological context, i.e. combined with the
proportionality $M\propto N^{-1/9}$ define 
\be
M^{10}\int_{M_{10}} e^{\wedge 10}= N^2 N^{-10/9} =N^{8/9}
\label{inte}
\ee
and the possible error one makes (due to different N dependence in the 
M.D. phase) is of order $N_0/N_{eq}\sim 10^6$ at most (but probably smaller).
This is indeed like fixing the value of the integral at an initial 
time to be of order one and let it be defined by time evolution afterwards.

Now let us couple this result with the assumption that still $g^2N=1$, and 
then the spatial integral of the cosmological constant term
(with a $1/g^2 N$ put in for free) is 
\be
(\frac{1}{g^2N}) \frac{1}{M} M^{11}\int _{M_{10}} e^{\wedge 10}=
(\int d^3 x e^{\wedge 3})(\frac{\Lambda}{M_P^2})\frac{M_P^4}{M} = N^{8/9}
\label{integ}
\ee
(the last equality comes from (\ref{inte}))
and substitute the spatial volume of the horizon and get 
\be
(L_0)^3\frac{\Lambda}{M_P^2}\frac{M_P^4}{M} \sim N^{8/9}
\ee
an equality which is satisfied to a remarcable degree. The l.h.s is 
\be
(10^{42} GeV^{-1})^3 10^{-123} (10^{19}GeV)^4 \frac{1}{M}=\frac{10^{79}}{
M[GeV]}
\ee
which is $10^{78}$ when substituting the value of M=10 GeV, whereas the 
r.h.s. is approximately $10^{77}$! Incidentally, (\ref{integ}) looks now like 
(\ref{alpha}), with $\alpha=2-1/9$, close to what one would have expected as a
natural definition ($\alpha=2$).

Let's remark now that we could have defined maybe $g^2N^2=1$ (as we 
mentioned, there is no constraint $g^2N =1$), in which 
case $M_{P,11}=MN^{2/9}$, and M stays the same ($\sim 10 GeV$), 
since $\Lambda/M^2 
=1/N$, and $\Lambda$ is determined from experiment. Then $N^{2/9}\simeq 
10^{19}$, and hence the fundamental parameter $M_{P,11}\simeq 10^{20} GeV
\simeq 10 M_{P,4}$, that is, of the order of the Planck scale! 
That means that the radii of the internal dimensions are 
also of Planck scale. This is remarcable, but it seems to work worse 
than before. Then however 
\be
M^{10}\int_{M_{10}} e^{\wedge 10}= N^2 N^{-20/9}=N^{-2/9}
\ee
so M is now almost the inverse scale of the Universe (in which case the 
r.h.s. would have been 1) and 
\be
\frac{1}{g^2N} \frac{1}{M} M^{11}\int _{M_{10}} e^{\wedge 10}=
(\int d^3 x e^{\wedge 3})\frac{\Lambda}{M_P^2}\frac{M_P^4}{M} = N^{7/9}
\ee
which seems to be a worse match. But if one remembers that the volume 
could be underestimated by a factor $N_0/N_{eq}\simeq 10^6$, maybe the 
r.h.s. is $10^6 \times N^{7/9}$, which is again close enough to the l.h.s. 
The cosmology one gets now is unfortunately also worse.

So let us say what kind of cosmology one expects. We have seen that the order 
of magnitude of the cosmological constant today is predicted. But $\Lambda$
depends on the number of particles inside the horizon, which varies. More 
precisely, in the first case ($g^2N=1$),
\be
\Lambda\propto M^2 N^{-1}\propto N^{-11/9}
\ee
But we saw that $N\propto t $ in the M.D. era and $N\propto t^{3/2}$ in the
R.D. era, so that 
\be
\rho_{\Lambda}=\Lambda\propto t^{-11/9} (M.D.)\;\; t^{-11/6} (R.D.)
\ee
Since in both cases the decrease is slower than for the leading component,
(which goes like $1/t^2$ in both  cases) 
and $\Lambda$ becomes dominant just now, it was subleading in the past. 
The density of a subleading component is 
\be
\rho \sim R^{-3(1+w)}\sim t^{-2(1+w)} (M.D.) 
\ee
and 
\be
\rho \sim R^{-3(1+w)}\sim t^{-\frac{3}{2}(1+w)} (R.D.) 
\ee
which means that 
\be
w_{\Lambda}(M.D.)=-7/18\;\; w_{\Lambda}(R.D.)=2/9
\ee
A separate problem is whether one can associate such a simple model (effective
w based on just the evolution of the varying $\Lambda$) with experimental 
constraints. It is not clear how to treat correctly a time varying $\Lambda$
in this scenario, where $\Lambda$ is determined by N, so we will stick to the 
effective w even though its justification is lacking.

This seems to be marginally compatible with observations, which are mainly in 
the M.D. era and strongly support an accelerating universe 
($-q=a\ddot{a}/(\dot{a})^2, q_0=\Omega_0(1+3w)/2$), that is $w\leq -1/3$, 
but besides that support a time dependent cosmological constant under certain 
conditions. 

Also note that today, if $\Lambda$ becomes dominant, the effective w of the 
universe is probably negative. Note that since 
\be
R(t)\sim t^{\frac{2}{3(1+w)}}
\ee
and since the number of particles in the horizon goes like 
\be
N\sim t^3/R(t)^3 \sim t^{\frac{1+3w}{1+w}}
\ee
then if the effective w is close to -1/3, N is approximately constant, so 
$\Lambda$ is too! Of course the real cosmology is hard to describe. 
In particular, there is no selfconsistent solution to having an effective 
fluid generated by $\Lambda$ alone, such that 
\be
\rho_{\Lambda}=\Lambda\sim N^{-11/9}\sim t^{-\frac{11(1+3w)}{9(1+w)}}
\sim t^{-3(1+w)}
\ee
but what is sure is that the effective w of the Universe is smaller than 
-7/18 (when we put the effective w =-1/3, we get $\Lambda$=ct. which has 
w=-1, i.e. smaller).

In the second case, if $g^2N^2=1$, we saw that $\Lambda 
\sim N^{-13/9}$, which would mean 
\be
\rho_{\Lambda}=\Lambda\propto t^{-13/9} (M.D.)\;\; t^{-13/6} (R.D.)
\ee
One sees imediately the problem, since in R.D. era, $\rho_r\sim t^{-2}$, 
so $\Lambda$ decays faster than the dominant energy, so it would 
have dominated as some point in the (not so distant, i.e. 
not at the inflation time, but very close to now) past. Also, now
we would have $w_{\Lambda}(M.D.)=-5/18$, which means the universe 
is not accelerating.

Let us now analyze the particle physics consequences of these scenarios 
and the possible relation to M theory. 
First, if N really represents the number of particles in the system, 
then in experiments other than cosmology N is not the number of particles 
in the horizon, but rather the number of particles in the system. 
Let us assume that $M_{P,11}$ is constant. Then on 
Earth, we could experiment at most with the number of particles in Earth.
The Earth mass is about $3 \times 10^{51}GeV$, and Earth is mostly C, Si, O, 
etc., things with 10-30 nucleons (each about 1 GeV), so let's say
$N_E\sim 10^{50}$, with $M=M_{P, 11}N^{-1/9}= 10^{7.5} TeV \times 10^{-5.5}
\sim 100 TeV$, and with smaller numbers even higher. All of these are 
consistent with observations! (new physics at about 1-10TeV). So even if 
$M_{cosmo}\sim 10 GeV$ (which may still be wrong by some orders of magnitude),
still one is not contradicting Earth based experiments! And for that matter 
one isn't contradicting cosmology either, since for 
a temperature $T\sim 10 GeV$, compared 
to the temperature of the equality of matter and radiation energies
$T_{eq}\sim 5 eV$, N changes as $N_{HOR}\sim t^{3/2}\sim R^3 \sim T^{-3}$, 
that is changes by a factor of $10^{9\times 3}$. After equality 
until now we have a change of $10^6$, for a total of $10^{33}$, that is 
$N\sim 10^{54}$, which means that M has now been shifted to 10 TeV, almost 
like before. 

A puzzling fact is that 
 the number of particles in the horizon at the Planck scale 
is of order 1, by extrapolating the R.D. result, since $N_{HOR}\sim 
T^{-3}$, and at 3 GeV one has $N_{HOR}=10^{54}$, that means $N_{HOR}=1$ 
at $ 3\times 10^{18} GeV$. This is odd, since $M_{P,4}$ is a derived 
quantity, and $M_{P,11}$ is the fundamental scale, so there would be 
no reason to fix $N_{HOR}=1$ at $M_{P,4}$.  But presumably 
the behaviour of $N_{HOR}$ with the temperaure T changes drastically 
when the temperature T is at the internal size scale, $R^{-1}
=10^{4} TeV$, such that until the temperature gets
 to $10^{7}TeV$ one actually has 
$N_{HOR}=1$. If nothing happens up to 3 TeV, then one has $N_{HOR}=10^{45}
$, which means one needs an average behaviour of $N_{HOR} \sim T^{-6.5}$ 
between the two scales. 

At this point however one can ask can we relate this theory 
(low energy limit of CS) to M 
theory? The problem is that if $1/g^2=N$, there are two energy scales 
in the theory: the scale $M_{P,11}= 10^{7.5} TeV$ and the scale M=10GeV, which 
varies with N if $M_{P,11}$ is to be fixed. That is hard to understand in 
M theory, all the more since one doesn't have 1/N corrections as in usual M 
theory (with N=number of particles in the system), but rather $M^{-2}=
(M_{P,11})^{-2}N^{2/9}$ corrections! So it is unclear whether there is a 
 relation to M theory.

But there seems to be a way out, in assuming that $1/g^2=1$ instead
(case 3 and final case to be analyzed). 
Assume then that the quantized coupling $1/g^2$ relates superselection 
sectors and that we are looking at the sector with $1/g^2=1$. Then 
\be
M_{P,11}=M=10\; GeV
\ee
(which one can assume is modified a bit by uncertainties, hopefully up 
to $\sim 10TeV$), and consequently the average radius of the internal 
directions is given by 
\be
\frac{M}{R^{-1}}\sim (\frac{M_{P,4}}{M})^{2/7}
\ee
and if $M\sim TeV$, then $R^{-1}\sim M 10^{-4.5}\sim 30 MeV$.
Then $M, M_{P,11}, R$ are independent of N, and since $\Lambda/M^2=1/N$, 
$\Lambda \propto N_{HOR}^{-1}$. Also then 
\be
M^{10}\int _{M_{10}} e^{\wedge 10}\propto N^2
\ee
and so one can fix 
\be
(\frac{1}{g^2N}) \frac{1}{M} M^{11}\int _{M_{10}} e^{\wedge 10}=
(\int d^3 x e^{\wedge 3})\frac{\Lambda}{M_P^2}\frac{M_P^4}{M} = N
\ee
which would imply that $\int_{M_{10}}J =N^2$ (which we said is 
the most natural), and then also 
\be
(L_0)^3\frac{\Lambda}{M_P^2}\frac{M_P^4}{M} = N
\ee
which is still well satisfied experimentally. 
Note though that it is a bit unclear why this should be equal, since 
when the temperature  reaches $R^{-1}$, $N_{HOR}$ is still huge, and 
presumably the volume depends differently on $N_{HOR}$ after that, so 
it might be a more complicated formula on the r.h.s. of the  previous 
equation. In any case, even if we assume the r.h.s. is exactly N, we come to a 
reasonable enough agreement with data, since as we saw, the l.h.s. is 
$=10^{79}/M[GeV]$.

Let's see what cosmology this gives. Since $\Lambda\propto N_{HOR}^{-1}$,
\be
\rho_{\Lambda}=\Lambda\sim t^{-1} (M.D.)\;\;\; t^{-3/2} (R.D.)
\ee
and so effectively 
\be
w_{\Lambda}(M.D.)=-1/2, \;\; w_{\Lambda}(R.D.)=0
\ee
which is even better compared with experiment. 

\section{Conclusions}

In this paper I have analyzed the possible relation of M theory with CS 
supergravities. Based on the approach of Horava in \cite{horava} I analyzed 
the high energy limit of a CS action. The $OSp(1|32)\times OSp(1|32)$ 
supergroup contracts to the D'Auria-Fre supergroup, with a mismatch in 
numbers, which can be attributed to the 0-form trick used by the latter. 
The action which I propose is obtained in a spinor representation for 
the supergravity fields, where the group generators are gamma matrices. 

The theory has a covariant formulation in (10,2) dimensions, without 
being explicitly a gravitational (10,2) theory. One has only a (10,2) 
spin connection, so one can introduce a fake vielbein satisfying the 
vielbein postulate $De^A=0$. It would be interesting to see whether this 
formulation can be related to the formulation with unit determinant on the 
compact space of \cite{tse}. The CS supergroup contracts to the IIB algebra, 
but only via the usual T duality, extended with an extra freedom due to the 
many new fields in the theory. One could dimensionally reduce also 
to a (9,2) dimensional theory, by just reinterpreting the fields in (10,1) 
dimensions (mixing up the spin connection and vielbein components, for 
instance). The prototype for this, the 3d gravity, is too simple in a 
sense: the reinterpretation effectively changes space with time ((2,1) 
vs. (1,2) signature), but in higher dimensions it is harder. Whether the 
rewriting of the 11d CS theory as (9,2) is entirely consistent and what are
its consequences deserves further study.

The equations of motion of the 11d CS supergravity were studied in the high 
energy limit. They are solved by the equations of motion of 11d supergravity,
linearized in everything but the vielbein. It is not clear whether this is 
the unique solution, but it is a solution. A possible caveat here is that 
we needed to introduce extra constraints on $A_{(3)}$, but these were related
to the presence of 11d sugra interaction terms. The high energy limit was then 
further studied using the assumption of Horava that one can introduce CS 
matter (via Wilson lines) and average, having in effect a cosmological 
constant term. 

The fact that both the linearized equations of motion (with 
first order gravity -the vielbein- still nonlinear!) and the invariance 
supergroup of usual 11d sugra are obtained means that one should obtain the 
full nonlinear sugra somehow. The exact mechanism seems obscure at the moment.
The difficulty is partly to realize how to couple matter. The usual CS matter 
(Wilson line) seems an ideal candidate for point particle coupling, 
since it comes with the minimum 
momentum in the contraction limit, M. However, it is only linear, whereas the 
equations of motion are 5th order in R. This might be only a sign of our lack 
of a full description of a consistent theory. 

The observational consequences of the theory were also analyzed. The 
constraints on the observed $\Lambda$, namely $\Lambda\sim 10^{-123}M_{P,4}^4$
imply that the value of M is constrained by $\Lambda\sim M^2 M_{P,4}^2$, to 
be $M\sim 1/L_0$ ($L_0$ is the size of the horizon today). Reversing the logic 
and imposing that M is of the order of $1/L_0$ we obtain a prediction for 
the cosmological constant! For the identification of the parameter N, one has 
to make assumptions about what the partons are. The most natural assumption in 
the context of M theory is that the partons are D0 branes, although the 
concrete realization of that idea is still lacking. Coupled with the idea 
that the horizon gives the size of the system one gets $N\sim N_{HOR}^{1/2}$.
Consequently, $M_{P,11}=MN \sim 10 GeV$, but presumably this can be driven 
up to 1TeV or higher. Here note that if one takes the point of view that the 
size of the system is important for the determination of $M_{P,11}$ in 
each experiment, $M_{P,11}$ could be driven still up (by having M higher 
and N smaller). The most naive assumption, that the cosmological constant 
behaves as $M^2$, 
and M as $1/L_0\sim 1/t$, implies that the cosmological constant 
looks like a dominant matter component in FRW Universe. Moreover, $M_{P,11}$
varies too much in time. On top of that, a CS constraint on the size of the 
universe seems hard to satisfy. The cosmological consequences of such a model 
deserve further study, it seems quite rich in possible phenomena.

The role of M and N in M theory seems to be that of extra parameters 
of M theory determining the cosmological model, obtaining the usual M theory 
in the high energy limit. If one views $M_{P,11}$ as fundamental, then 
M corrections are $M_{P,11}/N$ corrections. So somehow the definition of 
M theory as a CS theory with Wilson line matter should reproduce usual M 
theory at high energy. 

I have analyzed the possibility for a quantum theory
 of such CS models, and found it hard to 
understand. Certainly one can't use the usual perturbative expansion. 
Moreover, there seem to be contradictions with the known quantum corrections 
of string and M theory (namely that these are quantum corrections which are
not of CS form, and that there are no $R^2$ and $R^3$ corrections in M 
theory), but first of all these are conclusions derived from 
usual perturbation theory. Secondly, the inclusion of nonlinear 11d sugra
matter terms in the CS theory is already unclear. The remaining hope is that
the M expansion obsures the CS origin of both the 11d sugra and its quantum 
corrections.

Finally, the possibility of having a CS gravity theory in the {\em low 
energy} expansion was analyzed. In the CS equations of motion, 
one gets a large 
mass term for the 3-form. But if one truncates just to the 
gravitational sector, 
one might have a chance at a good phenomenology. Now the only assumption 
consistent with observations is that the partons are particles, not D0 
branes (strictly speaking, we could still have D0 branes, and the assumption 
that the mismatch in cosmological constant is of order $1/N^2= 1/N_{HOR}$.
That possibility has not been considered, but it gives worse matches 
with experiment).
Depending on the relation of N to the quantized CS coupling $k=1/g^2$, we 
get different cosmologies: 1) k=1. Then $M_{P,11}= M\sim 10GeV$, and 
$w_{\Lambda}(M.D.)=-1/2, w_{\Lambda}(R.D.)=0$, 
(here note that it is not clear that the effective w makes sense. $\Lambda(t)$
comes from N(t), and it is not clear how to describe this model consistently)
and $\int _{M_{10}}J=N^2$, 
which is close to the experimental data. 2) k=N. $M_{P,11}\sim 10^{7.5}
GeV$ and $w_{\Lambda}(M.D.)=-7/18, w_{\Lambda}(R.D.)=2/9$, and $\int 
_{M_{10}}J=N^{8/9}$, which is remarcably well satisfied experimentally. 
3)$k=N^2$. $M_{P,11}\sim 10M_{P,4}$! (only one gravity scale). Then however
the cosmological constant would have dominated in the past, since $\Lambda
(R.D.) \propto t^{-13/6}$, and also $w_{\Lambda}(M.D.)=-5/18>-1/3$, so the 
Universe would be decelerating. The constraint $\int _{M_{10}}J= N^{7/9}$ 
would be also less well satisfied.

In conclusion, 
one can say that we have just scratched the surface of the possible relation 
of CS supergravities with M theory, and there are many things left to do.

{\bf Ackhowledgements} I would like to thank Peter van Nieuwenhuizen for 
getting me familiarized with the problem and with the work of D'Auria-Fre. 
I would also like to thank Steve Corley for many insightful discussions
on these issues and Amihay Hanany and Sanjaye Ramgoolam for discussions.

\newpage

{\Large\bf{Appendix A. CS group invariant}}
\renewcommand{\theequation}{A.\arabic{equation}}
\setcounter{equation}{0}

Conventions: Throughout the paper, I have used A,B,C,... for 11d 
tangent indices, $\Pi,\Sigma,\Omega,...$ for (10,2) tangent indices, and 
a,b,c,... for 10 and lower (and for general dimension) tangent indices. 
Curved indices were indiscriminately denoted by $\mu, \nu, \rho,...$ since they
appear more rarely. M,N,P,... denote group indices in the corresponding 
representation. In section 5, m,n,p,... represent 9d tangent indices. 

Let us 
calculate more precisely the group invariant. As we saw, the 11d decomposition
of 12 gamma matrices is 
\be
\Gamma_{AB}=\gamma_{AB} \otimes 1, \;\; \Gamma_{A 12}=\gamma_{A} 
\otimes (-\sigma _3), \;\; \Gamma_{A_1...A_5} =
\gamma_{A_1...A_5}\otimes \sigma _1
\ee
which means that in order to get a nontrivial result for the trace 
$Tr [ (\Gamma^{(i)})^6\Gamma_{13}]$ the number n of $Z^{(5)}$s is 0 mod 2, 
the number m of $Z^{(1)}$s is 1 mod 2, and $Z^{(2)}$s p=6-n-m. It is easy 
to see that (n,m,p)= (0,1,5) is the usual term and (0,3,3) and (0,5,1) 
give zero, having 9 and 7 indices respectively. So we need to calculate
the 11d traces $Tr[(\gamma^{(5)})^2(\gamma^{(2)})^3\gamma^{(1)}], 
Tr[(\gamma^{(5)})^2\gamma^{(2)}(\gamma^{(1)})^3]$ and 
$Tr[ (\gamma^{(5)})^4\gamma^{(2)}\gamma^{(1)}]$. Also note that one 
only needs the part symmetrized under the exchange of curvatures of the 
same type. 

Therefore let's calculate
\be
T_1= \frac{1}{32} 
Tr [ \frac{1}{2} \{ \gamma^{A_1...A_5}, \gamma^{B_1...B_5} \} 
\frac{1}{2}\{ \frac{1}{2}\{ \gamma^{C_1C_2}, \gamma^{C_3C_4} \} , 
 \gamma^{C_5C_6} \} \gamma^{C_7} ]
\ee
(normalized by the trace of the identity) and 
\be
T_2= \frac{1}{32} Tr [ \frac{1}{2} \{
\frac{1}{2} \{ \gamma^{A_1...A_5}, \gamma^{B_1...B_5} \} ,
\frac{1}{2} \{ \gamma^{D_1...D_5}, \gamma^{E_1...E_5} \} \} \gamma^{C_1C_2}
\gamma^{C_7} ]
\ee
If there is a further symmetrization to be obtained we will assume it 
implicitly.
Let us first establish a few gamma matrix lemmae. If $A_1,...A_{k+1}, B_1,...
B_n, C_1,...C_m$ are all different, then we have (here we don't have any 
summation over repeated indices) 
\bea
&&\gamma^{A_1...A_{k+1}B_1...B_n}\gamma^{A_1...A_{k+1}C_1...C_m}
= (-)^{\frac{(k+1)}{2}(k+2n)}\gamma^{B_1...B_nC_1...C_m}
\nonumber\\&&
= (-)^{(m-n)(k+1)+m n }
\gamma^{A_1...A_{k+1}C_1...C_m}\gamma^{A_1...A_{k+1}B_1...B_n}
\eea
and therefore 
\bea
&&\gamma^{B_1...B_n}\gamma^{C_1...C_m}
=\sum_k (k+1)! \times \begin{pmatrix}
n&\\k+1& \end{pmatrix} \times \begin{pmatrix} m&\\k+1 & \end{pmatrix}
\nonumber\\&&
\times (-)^{\frac{k+1}{2}(2n-k-2)}\delta^{[B_1...B_{k+1}}_{[C_1...C_{k+1}}
{\gamma^{B_{k+2}...B_n]}}_{C_{k+2}...C_m]}
\eea
Here I have defined the delta symbols with strength one, that is 
\be
\delta^{[A_1...A_n]}_{[B_1...B_n]}= \frac{1}{n!}(\delta^{A_1}_{B_1}...
\delta^{A_n}_{B_n}+ (n!-1) terms)
\ee
so that 
\bea
&&\delta_{A_1...A_n}^{B_1...B_n}M_{B_1...B_n}=M_{A_1...A_n}\nonumber\\
&&\epsilon^{A_1...A_nC_1...C_{11-n}}\epsilon_{B_1...B_nC_1...C_{11-n}}
=-n!(11-n)!\delta_{B_1...B_n}^{A_1...A_n}\nonumber\\
&&
\gamma^{A_1...A_n}=\epsilon^{A_1..A_nB_{k+1}...B_{11}}\gamma_{B_{k+1}...B_{11}}
\frac{(-)^{[k/2+1]}}{(11-k)!}
\eea
Then one has
\bea
&&\frac{1}{2} \{ \gamma^{AB}, \gamma^{CD} \} =\gamma^{ABCD} -2 \delta^{AB}_{CD}
\\&&
\frac{1}{2} \{ \gamma^{A_1...A_5}, \gamma^{B_1...B_5} \} = 25 (
\delta^{[A_1}_{[B_1}{\gamma^{A_2...A_5]}}_{B_2...B_5]}-24 \delta^{[A_1A_2A_3
}_{[B_1B_2B_3}{\gamma^{A_4A_5]}}_{B_4B_5]}+\frac{24}{5}
\delta^{A1...A_5}_{B_1...B_5} )\label{fourgamma}\\
&& \frac{1}{2} \{ \gamma^{ABCD} , \gamma^{EF} \} = 
\gamma^{ABCDEF}-12 \delta^{[AB}_{EF}\gamma^{CD]}
\eea
and then 
\bea
&&T_1=-50 [12 \delta^{[A_1A_2A_3} _{[B_1B_2B_3} {\epsilon^{A_4A_5]}}_{B_4B_5
]C_1...C_7}+\delta^{[A_1}_{[B_1}{{\epsilon^{A_2...A_5]}}_{B_2...B_5]}}
^{C_7C_5C6}\delta^{C_1C_2}_{C_3C_4}]\nonumber\\&&
+ 6 \delta^{[A_1}_{[B_1}{{\epsilon^{A_2...A_5]}}_{B_2...B_5]}}
^{C_7[C_3C_4}\delta^{C_1C_2]}_{C_5C_6}
\eea
Using (\ref{fourgamma}) and then computing the anticomutator with itself
and not writing explicitly the obvious antisymmetrization ($A_1...A_5; 
B_1...B_5, D_1...D_5, E_1...E_5$), we get 
\bea
&&T_2= (25)^2 48 [ 12 \delta^{B_1B_2B_3}_{A_1A_2A_3}\delta^{D_1D_2D_3}
_{E_1E_2E_3}{{\epsilon^{B_4B_5D_4D_5}}_{A_4A_5E_4E_5}}^{C_1C_2C_7}
\nonumber\\
&& +\frac{1}{5} \delta^{A_1}_{B_1}{{\epsilon^{A_2...A_5}}_{B_2...B_5}}
^{C_1C_2C_7}
\delta_{E_1...E_5}^{D_1...D_5}
+36\delta^{A_1}_{B_1}{\epsilon^{A_2...A_5}}_{B_2...B_5 FGH}
\delta^{D_1D_2D_3}_{E_1E_2E_3}\delta ^{FG}_{[D_4D_5}\delta^{C_1C_2C_7}
_{E_4E_5]H} ]\nonumber\\&&
=(25)^2 48 [ 12 \delta^{B_1B_2B_3}_{A_1A_2A_3}\delta^{D_1D_2D_3}
_{E_1E_2E_3}{{\epsilon^{B_4B_5D_4D_5}}_{A_4A_5E_4E_5}}^{C_1C_2C_7}
\nonumber\\
&& +\frac{1}{5} \delta^{A_1}_{B_1}{{\epsilon^{A_2...A_5}}_{B_2...B_5}}
^{C_1C_2C_7}
\delta_{E_1...E_5}^{D_1...D_5}
+7\cdot 48\delta^{A_1}_{B_1}{\epsilon^{[A_2...A_5B_2B_3}}_{C_1C_2C_7[D_4D_5}
\delta^{D_1D_2D_3}_{E_1E_2E_3}\delta ^{B_4B_5]}_{E_4E_5]}
\eea
(where in the last line the trace was calculated in a different way)

\end{document}